\documentclass[11pt,a4paper]{article}
\usepackage{jheppub}
\usepackage{amsmath,amsthm,amssymb}
\usepackage[dvipsnames,svgnames,x11names]{xcolor}
\usepackage[mathscr]{eucal}
\usepackage{braket}
\usepackage{bm}
\usepackage{mathtools}
\usepackage[utf8]{inputenc}
\usepackage[T1]{fontenc}
\usepackage{hyperref}
\hypersetup{
    colorlinks=true,                 % false: boxed links; true: colored links
    linkcolor=blue,%SeaGreen,                 % color of internal links (change box color with linkbordercolor)
    linktoc=all,                     % link over complete toc entries
    citecolor=blue,                 % color of links to bibliography
    filecolor=blue,                  % color of file links
    urlcolor=blue                 % color of external links
}
\usepackage{url}
\theoremstyle{plain}
\newtheorem{theorem}{Theorem}
\numberwithin{theorem}{section}

\theoremstyle{definition}
\newtheorem{definition}[theorem]{Definition}
\newtheorem{conjecture}[theorem]{Conjecture}

\newtheorem*{theorem*}{Theorem}
\newcommand{\defeq}{\mathrel{\mathop:}=}

\DeclareMathOperator{\Tr}{Tr}

\DeclareMathOperator{\supp}{supp}

\newcommand{\Rmnum}[1]{\expandafter\@slowromancap\romannumeral #1@}

\newcommand{\be}{\begin{equation}}
\newcommand{\ee}{\end{equation}}

\renewcommand{\a}{\alpha}	%%% Redefinition
\newcommand{\cH}{\mathcal{H}}

\newcommand{\esc}{\mathrm{esc}}
\newcommand{\off}{\mathrm{off}}
\newcommand{\one}{\mathbf 1}
\newcommand{\dd}{\,\mathrm d}

\newcommand{\az}{$\a$--$z$}

\title{Quasi-local form for $\alpha$--$z$ R\'enyi QNEC\\ from fixed-ray escorts}
\author[a]{Tanay Kibe}
\emailAdd{tanay.kibe@ib.edu.ar}
\affiliation[a]{Instituto Balseiro, Centro 
At{\'o}mico Bariloche, S.C. de Bariloche, 8400, 
R{\'i}o Negro, Argentina}
\author[b]{and Pratik Roy}
\emailAdd{roy.pratik92@gmail.com}
\affiliation[b]{Institute of Mathematics, University of Warsaw, ul. Banacha 2, 02-097 Warsaw, Poland}
\begin{document}

\abstract{
R\'enyi divergences generalize relative entropy and measure the distinguishability of quantum states,
but their relation to energy and entanglement entropy is less direct
than for ordinary relative entropy. 
We introduce and study an energy--entropy relation for R\'enyi divergences 
by making use of an integral representation that expresses the $\alpha$--$z$
R\'enyi divergence as an average of ordinary relative entropies of
auxiliary escort states constructed from the excited state and the
reference state. The integral representation gives energy--entropy
decompositions, averaged entanglement laws and energy bounds in
regulated quantum field theory (QFT).
We further obtain a quasi-local expression for the
diagonal second null shape variation of the $\alpha$--$z$ R\'enyi
divergence of an excited state relative to the QFT vacuum.  This expression contains a
transport term because forming an escort after restricting to a
smaller region generally differs from restricting an escort formed
beforehand. The conjectural $\alpha$--$z$ R\'enyi quantum null energy 
condition (QNEC) then gives a lower bound on the
null energy density of the average escort state. As in the ordinary
QNEC, the bound involves a second null variation of von Neumann
entropy, here averaged over the escort ensemble.
We verify the conjecture for an exactly solvable family
of coherent excitations of a free scalar field throughout
the finite-divergence part of the upper data-processing region.
For this
family, the sandwiched R\'enyi divergence at fixed $\alpha>1$ has a
positive diagonal second null shape variation that is entirely
accounted for by the transport term.
We also show that escort formation commutes with the reduced encoding
channel in finite-dimensional operator-algebra quantum error-correcting
codes with exact complementary recovery.
}

\maketitle

\newpage

\section{Introduction}

Relative entropy relates the distinguishability of quantum states to
energy and entanglement. To see this, consider a local quantum field theory, and let $\rho$ and $\sigma$ be the
density matrices of an excited state and the vacuum restricted to a
region, with a common ultraviolet regulator understood. Their relative
entropy is
\begin{equation}
 D(\rho\Vert\sigma)
 =\operatorname{Tr}\!\left[(\rho-\sigma)K_\sigma\right]
  -\bigl[S(\rho)-S(\sigma)\bigr],
 \qquad K_\sigma=-\log\sigma.
 \label{eq:intro-relative-free-energy}
\end{equation}
Here the first term is the change in the vacuum modular energy, and the
second is the change in entanglement entropy. When the modular
Hamiltonian has a local expression in terms of the stress tensor, this
identity gives a direct connection between distinguishability and
energy. The purpose of this paper is to find a corresponding
interpretation for R\'enyi divergences and to understand how it changes
when the region is varied.

A useful example is the quantum null energy condition (QNEC). Consider
regions bounded by cuts $x^+=V(y)$ of the null plane $x^-=0$, where $y$
labels the transverse coordinates. The states $\rho_V$ and $\sigma_V$
are restrictions of a fixed global excitation and the vacuum. Locality
of the vacuum modular Hamiltonian relates the diagonal second null
variation of relative entropy to the energy density at the cut,
\begin{equation}
 D''_{++}[V;y]
 =2\pi\langle T_{++}(V(y),0,y)\rangle_{\rho_V}
  -(\Delta S)''_{++}[V;y],
 \qquad
 \Delta S[V]=S(\rho_V)-S(\sigma_V).
 \label{eq:intro-qnec-relative-entropy}
\end{equation}
Here the double prime denotes the coefficient of the transverse delta
function in the second shape variation. The QNEC states that this
coefficient for relative entropy is nonnegative. Thus a convexity
property of distinguishability becomes a lower bound on local null
energy. The relative-entropy formulation of the QNEC also has an intrinsic meaning
in continuum QFT through Araki relative entropy
\cite{Araki:1976zv}, without assigning separate UV-divergent von Neumann entropies to the
Type~III local von Neumann algebras.

The QNEC was proposed as a nongravitational limit of the quantum
focusing conjecture \cite{Bousso:2015mna}. It has been established in
free theories \cite{Bousso:2015wca,Malik:2019dpg}, holographic theories
\cite{Koeller:2015qmn}, and general local Poincar\'e-invariant QFTs
\cite{Balakrishnan:2017bjg}, with operator-algebraic proofs developed in
\cite{Ceyhan:2018zfg,Hollands:2025glm,Faulkner:2026gcf}. Its applications include bounds
on nonequilibrium evolution
\cite{Mezei:2019sla,Kibe:2021qjy,Banerjee:2022dgv,Kibe:2024icu,Kibe:2025cqc}
and renormalization-group irreversibility in the presence of defects
\cite{Casini:2023kyj}.\footnote{Saturation and nonsaturation in
two-dimensional holographic CFTs are studied in \cite{Ecker:2019ocp}.
For a review of energy conditions in gravity, QFT and holography, see
\cite{Iizuka:2025xnd}.}
The information-theoretic approach to the QNEC \cite{Wall:2017blw}
suggests considering other measures of distinguishability that decrease
under quantum channels. This property is the data processing
inequality.

We consider the two-parameter family of $\alpha$--$z$ R\'enyi
divergences \cite{Audenaert:2015npv}. For normalized density matrices
with faithful $\sigma$, their finite-dimensional definition is
\begin{equation}
 D_{\alpha,z}(\rho\Vert\sigma)
 =\frac{1}{\alpha-1}
  \log\operatorname{Tr}\!\left[
   \left(
    \sigma^{\frac{1-\alpha}{2z}}
    \rho^{\frac{\alpha}{z}}
    \sigma^{\frac{1-\alpha}{2z}}
   \right)^z
  \right],
 \qquad \alpha,z>0,\quad \alpha\neq1.
 \label{eq:intro-alpha-z}
\end{equation}
The line $z=1$ gives the Petz divergence, while $z=\alpha$ gives the
sandwiched R\'enyi divergence $D_\alpha=D_{\alpha,\alpha}$
\cite{Muller-Lennert:2013liu,Wilde:2013bdg,Beigi:2013teh,Frank:2013rov}.
These divergences also have an intrinsic definition for normal states
on von Neumann algebras using Haagerup $L^p$ spaces
\cite{Kato:2023aro,Kato:2023hlj,Hiai:2024qve}.

Lashkari conjectured that $D_\alpha$, like relative entropy, has a
nonnegative diagonal second null variation \cite{Lashkari:2018nsl}.
For $\alpha>1$, this R\'enyi QNEC has been proved in free and
superrenormalizable bosonic and fermionic theories in dimensions $d>2$
\cite{Moosa:2020jwt,Roy:2022yzm}, for integer $\alpha\geq2$ \cite{Kibe:2026wsg} and more recently for general $\alpha>1$ \cite{Faulkner:2026yzn} in
QFTs with the relevant half-sided modular-inclusion structure. Counterexamples exist for $\alpha<1$
\cite{Moosa:2020jwt}. We consider the analogous diagonal convexity
conjecture for $D_{\alpha,z}$ \cite{Moosa:2020jwt} in its upper data-processing region,
$\alpha>1$ and $\max\{\alpha/2,\alpha-1\}\leq z\leq\alpha$
\cite{Zhang:2018roy,Hiai:2024qve}. We restrict to the diagonal null convexity statement, since a universal off-diagonal inequality fails for these
divergences \cite{Kibe:2026bcn}.

Even when null convexity of the $\alpha$--$z$ divergence is known, its energy--entropy interpretation
is not manifest from \eqref{eq:intro-alpha-z}. To obtain such an interpretation, we use the fixed-ray integral representation established in the
companion paper \cite{Kibe:2026dtz}. For a given $(\alpha,z)$, set
$c=z/\alpha$. Powers of $\rho$ and $\sigma$ define a family of
normalized auxiliary states $\rho_\beta^{(c)}$, called escorts, along
the ray $(\beta,c\beta)$. The family begins at
$\rho_1^{(c)}=\rho$. Under the support and finiteness assumptions
reviewed in Section~\ref{sec:finite},
\begin{equation}
 D_{\alpha,z}(\rho\Vert\sigma)
 =\int_{I_\alpha}
   D(\rho_\beta^{(c)}\Vert\sigma)\,\mathrm d\mu_\alpha(\beta),
 \qquad
 \mathrm d\mu_\alpha(\beta)
 =\frac{\alpha}{|\alpha-1|}\frac{\mathrm d\beta}{\beta^2},
 \label{eq:intro-integral-representation}
\end{equation}
where $I_\alpha$ is the interval with endpoints $1$ and $\alpha$, and the measure has total integral one. Thus the R\'enyi divergence is an
average of ordinary relative entropies, with the R\'enyi parameters carried
by the states in the integrand. This representation does not require
data processing and remains valid on general von Neumann algebras.
At $c=1$, its differential form is the refined sandwiched identity of
\cite{Bao:2019aol}. As $\alpha\to1$ at fixed $c$, it reduces to
ordinary relative entropy.

Applying \eqref{eq:intro-relative-free-energy} to each escort gives
\begin{equation}
 D_{\alpha,z}(\rho\Vert\sigma)
 =\operatorname{Tr}\!\left[
   (\overline\rho_{\alpha,z}-\sigma)K_\sigma
  \right]
  -\Delta S_{\alpha,z}^{\mathrm{esc}},
 \label{eq:intro-free-energy-decomposition}
\end{equation}
where $\overline\rho_{\alpha,z}$ is the average of
$\rho_\beta^{(c)}$ with respect to $\mu_\alpha$, and
$\Delta S_{\alpha,z}^{\mathrm{esc}}$ is the average of
$S(\rho_\beta^{(c)})-S(\sigma)$. The modular energy is therefore evaluated in
the average escort state, whereas the entropy is averaged over the
individual escorts.

The main new issue arises when the region changes. One can first form
an escort on a region and then restrict it to a smaller region.
Alternatively, one can restrict both $\rho$ and $\sigma$ first and
form the escort from the restricted pair. These operations generally
give different states. 
Consequently, the escorts associated with
successive cuts need not be restrictions of a fixed global state.

For the null cut $x^+=V(y)$ described above, 
differentiating the modular energy with respect to $V$ changes both the modular
Hamiltonian and the escort ensemble itself. The latter change must be
included in the energy--entropy relation.
Write $\eta_{\beta,c}[V]$ for the escort formed from
$(\rho_V,\sigma_V)$, and $\overline\eta_{\alpha,z}[V]$ for its
average. With
\[\Delta\mathcal S_{\alpha,z}^{\mathrm{esc}}[V]
=\int_{I_\alpha}[S(\eta_{\beta,c}[V])-S(\sigma_V)]
\,\mathrm d\mu_\alpha(\beta),\] we obtain
\begin{equation}
 (D_{\alpha,z})''_{++}[V;y]
 =2\pi\bigl\langle T_{++}(V(y),0,y)\bigr\rangle_
       {\overline\eta_{\alpha,z}[V]}
  -(\Delta\mathcal S_{\alpha,z}^{\mathrm{esc}})''_{++}[V;y]
  +\mathfrak X_{\alpha,z;++}[V;y].
 \label{eq:intro-local-rqnec}
\end{equation}
Here, the explicit variation of the vacuum modular Hamiltonian gives the
local stress tensor, and the remaining modular-energy response is the
escort-transport term $\mathfrak X_{\alpha,z;++}$, defined in
\eqref{eq:az-local-transport}. It accounts for the effect of
reconstructing the escorts as the cut moves.

Equation~\eqref{eq:intro-local-rqnec} is an exact identity under the ultraviolet
regulator and differentiability assumptions of
Section~\ref{sec:az-null-variations}. Following the R\'enyi QNEC for the sandwiched R\'enyi divergence ($\alpha=z$), the diagonal $\alpha$--$z$ R\'enyi QNEC conjecture requires that 
\[
    (D_{\alpha,z})''_{++}[V;y] \geq0,
\]
in parts of the data processing region where $D_{\alpha,z}$ is finite.
Using \eqref{eq:intro-local-rqnec}, this gives a lower bound on the escort-averaged null energy
in terms of the escort entropy variation and transport. 
We emphasize that this does not give a bound
on the energy of the original excitation, since there is no general
comparison between that energy and the escort-averaged energy.

As a concrete example, we consider a family of coherent excitations of the vacuum in a free
scalar field theory and compute their escorts explicitly as the null
cut is varied. Each escort is itself a coherent displacement of the
restricted vacuum, so its covariance and regulated entropy agree with
those of the restricted vacuum. The vacuum-subtracted escort entropy
therefore vanishes as the cut moves. For the particular profile we
study, the escort-averaged null energy density at the cut also
vanishes for the sandwiched R\'enyi divergence for every fixed $\alpha>1$. The
transport term then accounts for the entire positive diagonal second
null shape variation of the divergence.

For the same family we also evaluate the general $\alpha$--$z$
divergence and find a positive diagonal second null shape variation wherever the
divergence is finite. Finiteness requires $z>\alpha-1$, so the
calculation verifies the $\alpha$--$z$ R\'enyi QNEC conjecture throughout the finite-divergence
part of the upper data-processing region.

For a fixed region, the integral representation also gives an escort-averaged
first law of entanglement and Bekenstein-type bounds for vacuum balls
in conformal field theories. At second order around the reference
state, the Hessian is an average of Bogoliubov--Kubo--Mori forms
evaluated on the escort perturbations. This gives a new representation
of the known $\alpha$--$z$ information metric
\cite{May:2018tir,Wilde:2025nki}.

In finite-dimensional operator-algebra quantum error-correcting codes
with exact complementary recovery  used in models of holography \cite{Harlow:2016vwg}, escort
formation commutes with the reduced encoding channel. One can form
an escort from the bulk state and its reference and then encode it,
or first encode both states on the boundary region and then form
the escort there. The two procedures give the same boundary state.
The channel also maps the bulk escort average to the boundary
escort average and preserves the relative entropy of each escort
with respect to its reference state. The bulk--boundary equality
for the sandwiched R\'enyi divergence discussed in
\cite{Caginalp:2022uzd} thus has a counterpart at the level of the
individual escort states. This correspondence is exact in the code
model. A semiclassical bulk interpretation of these states requires
further assumptions.

The rest of this paper is organized as follows. Section~\ref{sec:finite} reviews the fixed-ray representation and
derives the energy--entropy decomposition.
Section~\ref{sec:qft} develops the entanglement laws, the vacuum-ball
bounds and the encoding result.
Section~\ref{sec:az-null-variations} derives the null shape variation identities and
states the diagonal convexity conjecture.
The coherent-state calculation is given in
Section~\ref{sec:null-coherent-transport}, and
Section~\ref{sec:conclusions} discusses open questions and future directions.
Appendix~\ref{app:frechet-derivatives} collects the operator
derivatives used in the calculations.

\section{Integral representation and modular-energy--entropy decomposition}
\label{sec:finite}

Throughout this section, \(\rho\) and \(\sigma\) are density matrices on a finite-dimensional Hilbert space \(\cH\).
We first review the fixed-ray integral representation of the $\alpha$--$z$
divergence from \cite{Kibe:2026dtz}, which extends the refined SRD identity
of \cite{Bao:2019aol}.  We then use it to derive exact modular-energy--entropy and Holevo decompositions analogous to the familiar identity for relative entropy.

\subsection{Integral representation}
For $\alpha,z>0$, $\alpha\neq1$, define \cite{Audenaert:2015npv}
\begin{align}
 Q_{\alpha,z}(\rho\Vert\sigma)
 &\defeq
 \Tr\!\left(
 \sigma^{\frac{1-\alpha}{2z}}
 \rho^{\frac{\alpha}{z}}
 \sigma^{\frac{1-\alpha}{2z}}
 \right)^z,
 \label{eq:az-moment}
 \\
 D_{\alpha,z}(\rho\Vert\sigma)
 &\defeq \frac{1}{\alpha-1}
 \log Q_{\alpha,z}(\rho\Vert\sigma).
 \label{eq:az-divergence}
\end{align}
If $\sigma$ is not faithful, let $P=\supp(\sigma)$ be its support projector. Positive powers of $\sigma$ are understood as usual by functional calculus. Negative powers of $\sigma$ are understood as inverse powers of the faithful restriction $\sigma_P\defeq\sigma\vert_{P\cH}$ on $P\cH$, extended by zero on the orthogonal complement $P^\perp\cH$. For $0<\alpha<1$, the expression
\eqref{eq:az-moment} is defined for arbitrary $\rho$ and
$\sigma$.  For $\alpha>1$, $\sigma^{(1-\alpha)/(2z)}$ is read as a generalized inverse on the support projection only
when $\supp(\rho)\le \supp(\sigma)$. We therefore set
\[
  Q_{\alpha,z}(\rho\|\sigma)
  =D_{\alpha,z}(\rho\Vert\sigma)
  =+\infty
  \qquad\text{if }\supp(\rho)\not\leq \supp(\sigma).
\]
If $\supp(\rho)\leq \supp(\sigma)$, all the quantities may equivalently be
computed in the support projection, where
$\sigma_P$ is faithful.  Matrices and states obtained in this subspace
will always be regarded as matrices on $\cH$ by extension by zero on
$P^\perp\cH$.

Define the characteristic ray through the endpoint $(\alpha,z)$ by
\begin{equation}
 c\defeq \frac{z}{\alpha}>0,
 \qquad (\beta,z_\beta)\defeq(\beta,c\beta).
 \label{eq:az-ray-slope}
\end{equation}
For $0<\alpha<1$, assume
$\supp\rho\leq\supp\sigma$.  For $\alpha>1$, assume that
$Q_{\alpha,z}(\rho\Vert\sigma)<\infty$; in this branch finiteness already
implies support inclusion.  In finite dimensions with faithful $\sigma>0$, the support assumption is automatic.

Assume that $\supp(\rho)\leq \supp(\sigma)$. For $\beta>0$, set
\begin{align}
 X_{\beta}^{(c)}(\rho\Vert\sigma)
 &\defeq
 \sigma^{\frac{1-\beta}{2c\beta}}
 \rho^{1/c}
 \sigma^{\frac{1-\beta}{2c\beta}},
 \label{eq:az-ray-factor}
 \\
 Q_{\beta}^{(c)}(\rho\Vert\sigma)
 &\defeq
 \Tr\!\left[(X_{\beta}^{(c)})^{c\beta}\right]
 =Q_{\beta,c\beta}(\rho\Vert\sigma),
 \label{eq:az-ray-moment}
\end{align}
and define the corresponding escort state by
\begin{equation}
 \rho_{\beta}^{(c)}
 \defeq
 \frac{(X_{\beta}^{(c)})^{c\beta}}
 {Q_{\beta}^{(c)}(\rho\Vert\sigma)},
 \label{eq:az-ray-escort}
\end{equation}
with
\begin{equation}
 \rho_{1}^{(c)}=\rho
 \label{eq:az-escort-at-one}
\end{equation}
for every $c>0$.  Under the support inclusion,
$Q_{\beta}^{(c)}(\rho\|\sigma)$ is finite and strictly positive for
every $\beta>0$, and the escort state
$\rho_\beta^{(c)}$ is well defined.

Define the closed intervals and measures
\begin{equation}\label{eq:intro-measure}
I_\alpha=
\begin{cases}
[1,\alpha],&\alpha>1,\\
[\alpha,1],&0<\alpha<1,
\end{cases}
\qquad
\dd\mu_\alpha(\beta)=
\begin{cases}
\displaystyle\frac{\alpha}{\alpha-1}\frac{\dd\beta}{\beta^2},&\alpha>1,\\[1.1ex]
\displaystyle\frac{\alpha}{1-\alpha}\frac{\dd\beta}{\beta^2},&0<\alpha<1,
\end{cases}
\end{equation}
where the measure \(\mu_\alpha\) is a probability measure.
The $\alpha$--$z$ divergence can be expressed as a fixed-ray average over the relative entropy of the escort states \cite{Kibe:2026dtz}
\begin{equation}
 D_{\alpha,z}(\rho\Vert\sigma)
 =\int_{I_\alpha}
 D\!\left(\rho_\beta^{(c)}\Vert\sigma\right)
 \dd\mu_\alpha(\beta),
 \qquad c=\frac{z}{\alpha}.
 \label{eq:az-fixed-ray-integral}
\end{equation}

The same statement holds intrinsically for normal states on
an arbitrary von Neumann algebra \cite{Kibe:2026dtz}. This operator-algebraic formula is the
regulator-independent statement that remains meaningful for Type~III local
algebras that arise in QFT.

If $0<\alpha<1$ and support inclusion is dropped, the pure average
\eqref{eq:az-fixed-ray-integral} must be replaced by a boundary-corrected
formula.  In finite dimensions, with $P=\supp\sigma$, put
\begin{equation}
 q_c\defeq
 \Tr\!\left[(P\rho^{1/c}P)^c\right].
 \label{eq:az-boundary-seed}
\end{equation}
If $q_c>0$, put
\begin{equation}
 \widehat\rho_c\defeq
 \frac{(P\rho^{1/c}P)^c}{q_c}
 \qquad\text{on }P\mathcal{H}.
 \label{eq:az-boundary-normalized-seed}
\end{equation}
Then
\begin{equation}
 D_{\alpha,c\alpha}(\rho\Vert\sigma)
 =\int_{I_\alpha}
 D\!\left((\widehat\rho_c)_\beta^{(c)}\Vert\sigma\right)
 \dd\mu_\alpha(\beta)
 -\frac{\alpha}{1-\alpha}\log q_c.
 \label{eq:az-boundary-corrected}
\end{equation}
If $q_c=0$, the divergence is infinite.  Since the regulated-QFT
applications below assume a faithful vacuum density matrix, this boundary
term will not be necessary for our purposes.

\subsection{Modular-energy--entropy and Holevo decompositions}
\label{sec:az-modular-decomposition}

The integral representation gives the $\alpha$--$z$ divergence an averaged
modular-energy--entropy interpretation.
Assume that $\sigma$ is faithful. Define the fixed-ray escort average state and the escort-averaged entropy by
\begin{align}
 \overline\rho_{\alpha,z}
 &\defeq
 \int_{I_\alpha}\rho_\beta^{(c)}
 \dd\mu_\alpha(\beta),
 \label{eq:az-escort-barycenter}
 \\
 S_{\alpha,z}^{\esc}(\rho\Vert\sigma)
 &\defeq
 \int_{I_\alpha}S(\rho_\beta^{(c)})
 \dd\mu_\alpha(\beta),
 \qquad
 S(\eta)\defeq-\Tr(\eta\log\eta).
 \label{eq:az-escort-entropy}
\end{align}
Here and below $c=z/\alpha$ is understood.  Let
\begin{equation}
 K_\sigma\defeq -\log\sigma
 \label{eq:az-modular-hamiltonian}
\end{equation}
be the reference modular Hamiltonian.  Applying
$D(\eta\Vert\sigma)=-S(\eta)+\Tr(\eta K_\sigma)$ to every integrand in
\eqref{eq:az-fixed-ray-integral} gives
\begin{align}
 D_{\alpha,z}(\rho\Vert\sigma)
 &=-\int_{I_\alpha}S(\rho_\beta^{(c)})
 \dd\mu_\alpha(\beta)
 +\Tr(\overline\rho_{\alpha,z}K_\sigma)
 \notag\\
 &=\Tr\!\left[(\overline\rho_{\alpha,z}-\sigma)K_\sigma\right]
 -\left[S_{\alpha,z}^{\esc}(\rho\Vert\sigma)-S(\sigma)\right].
 \label{eq:az-modular-free-energy}
\end{align}
Thus, with
\begin{align}
 \Delta_{\alpha,z}\langle K_\sigma\rangle
 &\defeq
 \Tr\!\left[(\overline\rho_{\alpha,z}-\sigma)K_\sigma\right],
 \\
 \Delta_{\alpha,z}S_{\esc}
 &\defeq
 S_{\alpha,z}^{\esc}(\rho\Vert\sigma)-S(\sigma),
\end{align}
one has the exact modular-energy--entropy decomposition of the $\alpha$--$z$ divergence
\begin{equation}
 D_{\alpha,z}(\rho\Vert\sigma)
 =\Delta_{\alpha,z}\langle K_\sigma\rangle
 -\Delta_{\alpha,z}S_{\esc}.
 \label{eq:az-modular-free-energy-boxed}
\end{equation}

There is a useful refinement of this relation.  
Define the fixed-ray escort Holevo
information
\begin{align}
 \chi_{\alpha,z}^{\esc}
 &\defeq
 S(\overline\rho_{\alpha,z})
 -S_{\alpha,z}^{\esc}(\rho\Vert\sigma)
 \label{eq:az-escort-holevo}
 \\
 &=\int_{I_\alpha}
 D\!\left(\rho_\beta^{(c)}
 \middle\Vert\overline\rho_{\alpha,z}\right)
 \dd\mu_\alpha(\beta)
 \geq0.
 \label{eq:az-escort-holevo-relative}
\end{align}
The second equality is the continuous-ensemble Donald--Holevo identity
\cite{Donald_1987},\cite[Lemma~4, Eq.~(16)]{Holevo:2004ots}. It also follows immediately by
expanding the relative entropy and using the definition of the escort average state. Concavity of the von Neumann entropy gives \(\chi_{\alpha,z}^{\esc}\geq0\).
Thus the Holevo term $\chi_{\alpha,z}^{\esc}$ is the mean distinguishability between the
individual fixed-ray escorts and their average.

Substituting
$S_{\alpha,z}^{\esc}=S(\overline\rho_{\alpha,z})-\chi_{\alpha,z}^{\esc}$
in \eqref{eq:az-modular-free-energy} yields
\begin{align}
 D_{\alpha,z}(\rho\Vert\sigma)
 =D(\overline\rho_{\alpha,z}\Vert\sigma)
 +\chi_{\alpha,z}^{\esc}.
 \label{eq:az-donald-holevo}
\end{align}
Equivalently,
\begin{equation}
 D_{\alpha,z}(\rho\Vert\sigma)
 =\Tr[(\overline\rho_{\alpha,z}-\sigma)K_\sigma]
 -\big(S(\overline\rho_{\alpha,z}) - S(\sigma)\big)
 +\chi_{\alpha,z}^{\esc}.
 \label{eq:az-three-term}
\end{equation}
The modular-energy term therefore depends only on the escort average state,
whereas the entropy term retains information about the full fixed-ray
escort ensemble.  The Holevo term measures precisely the information lost by
replacing that ensemble by its average.

Fix $c>0$ and take $\alpha\to1$ with $z=c\alpha$.  Then
$D_{\alpha,c\alpha}(\rho\Vert\sigma)\to D(\rho\Vert\sigma)$, and the
measure $\mu_\alpha$ concentrates at $\beta=1$,
so that \(\overline\rho_{\alpha,z}\to\rho\) and \(\chi_{\alpha,z}^{\esc}\to0\).  The two
forms above reduce to
\[
D(\rho\Vert\sigma)
=\Tr[(\rho-\sigma)K_\sigma]-\bigl(S(\rho)-S(\sigma)\bigr),
\]
as expected.

Equations \eqref{eq:az-modular-free-energy}--\eqref{eq:az-three-term} are
literal density-matrix identities and hence apply only to an appropriately regulated QFT.  In a
continuum QFT, the modular-energy and von Neumann entropy terms
are generally not separately defined.  Their intrinsic replacement is the
operator-algebraic relative-entropy integral from \cite{Kibe:2026dtz}.

\subsubsection{Sandwiched specialization}
\label{subsec:finite-sandwiched-slice}

The sandwiched R\'enyi divergence is obtained by restricting the
$\alpha$--$z$ family to $z=\alpha$, i.e., $c=1$. The escort
states take the form
\begin{equation}
 \rho_\beta^{\,\mathrm{s}}
 \defeq\rho_\beta^{(1)}
 =\frac{
 \left(
 \sigma^{\frac{1-\beta}{2\beta}}
 \rho\,
 \sigma^{\frac{1-\beta}{2\beta}}
 \right)^\beta}
 {Q_{\beta,\beta}(\rho\Vert\sigma)}.
 \label{eq:finite-sandwiched-escort-specialization}
\end{equation}
Under the support assumptions of \eqref{eq:az-fixed-ray-integral},
the sandwiched divergence $D_\alpha\defeq D_{\alpha,\alpha}$
therefore satisfies
\begin{equation}
 D_\alpha(\rho\Vert\sigma)
 =\int_{I_\alpha}
 D(\rho_\beta^{\,\mathrm{s}}\Vert\sigma)\,
 \dd\mu_\alpha(\beta).
 \label{eq:finite-sandwiched-integral-specialization}
\end{equation}
We denote the corresponding average state, average escort entropy,
and Holevo information by $\overline\rho_\alpha^{\,\mathrm{s}}$,
$S_\alpha^{\esc,\mathrm{s}}$, and $\chi_\alpha^{\esc,\mathrm{s}}$.
Their definitions are the same as those above restricted to $z=\alpha$. The energy--entropy and Holevo decompositions follow
directly from \eqref{eq:az-modular-free-energy} and
\eqref{eq:az-donald-holevo} with this choice.

\section{Escort-averaged entanglement laws}
\label{sec:qft}

Throughout Sections~\ref{sec:qft} and
\ref{sec:az-null-variations}, we work in a QFT with a fixed ultraviolet regulator such that reduced density matrices of QFT subregions and their von Neumann entropies are well defined. Regional states are thus
represented by density matrices, and powers, logarithms, entropies, and
their first and second variations are understood in the common regulator.
We freely use finite-dimensional matrix functional calculus and assume that all traces,
shape derivatives, and interchanges with the fixed-ray integral that
appear below are well defined. 
This section develops some of the physical content of the fixed-ray integral representation in this context.

Let \(A\) be a fixed spatial region in a regulated QFT, and let \(\rho_A\) be the reduced density matrix of an excited state and \(\sigma_A\) be that of the QFT vacuum. 

Consider a smooth normalized perturbation
\begin{equation}
 \rho_A(\varepsilon)
 =\sigma_A+\varepsilon X_A+\varepsilon^2Y_A
 +O(\varepsilon^3),
 \qquad
 \Tr X_A=\Tr Y_A=0.
 \label{eq:az-state-perturbation}
\end{equation}
Here $X_A$ and $Y_A$ are self-adjoint.  We assume throughout that
$\sigma_A$ is faithful and that the regulated state path remains in the
faithful state space for sufficiently small $\varepsilon$.  The regulator
is also assumed to make the expansions below uniform in
$\beta\in I_\alpha$, so that differentiation and the fixed-ray integral may
be interchanged.

For each $\beta\in I_\alpha$, let the escort states be
\begin{equation}
 \eta_{\beta,c;A}(\varepsilon)
 \defeq
 \bigl(\rho_A(\varepsilon)\bigr)_\beta^{(c)}
 =\sigma_A+\varepsilon\xi_{\beta,c;A}
 +\varepsilon^2\zeta_{\beta,c;A}
 +O(\varepsilon^3).
 \label{eq:az-escort-expansion}
\end{equation}
Normalization gives
$\Tr\xi_{\beta,c;A}=\Tr\zeta_{\beta,c;A}=0$.

Introduce the Fr\'echet derivative of the logarithm\footnote{Our notation and the
standard spectral formulas are reviewed in
Appendix~\ref{app:frechet-derivatives}, following
\cite{hiai2014introduction}.}:
\begin{equation}
 \mathcal T_{\sigma_A}(U)
 \defeq D(\log)_{\sigma_A}[U]
 =\int_0^\infty
 (\sigma_A+t\one)^{-1}U(\sigma_A+t\one)^{-1}\dd t,
 \label{eq:az-log-frechet}
\end{equation}
where $\one$ denotes the identity matrix.
For a normalized path
$\eta(\varepsilon)=\sigma_A+\varepsilon U+\varepsilon^2W+
O(\varepsilon^3)$, the entanglement entropy expansion is
\begin{equation}
 S(\eta(\varepsilon))
 =S(\sigma_A)
 +\varepsilon\Tr(UK_{\sigma_A})
 +\varepsilon^2\Tr(WK_{\sigma_A})
 -\frac{\varepsilon^2}{2}
 \Tr[U\mathcal T_{\sigma_A}(U)]
 +O(\varepsilon^3).
 \label{eq:az-entropy-expansion}
\end{equation}

\subsection{Escort-averaged R\'enyi first law}

Define
\begin{equation}
 \overline\eta_{\alpha,z;A}(\varepsilon)
 \defeq
 \int_{I_\alpha}\eta_{\beta,c;A}(\varepsilon)
 \dd\mu_\alpha(\beta).
 \label{eq:az-perturbative-barycenter}
\end{equation}
Integrating \eqref{eq:az-entropy-expansion} and comparing it with
$\Tr[(\overline\eta_{\alpha,z;A}-\sigma_A)K_{\sigma_A}]$ shows that all
terms linear in $\varepsilon$, as well as all terms involving the unknown
second-order escort coefficient $\zeta_{\beta,c;A}$, cancel in
\eqref{eq:az-modular-free-energy}.  Consequently,
\begin{equation}
 \left.\frac{\dd}{\dd\varepsilon}
 D_{\alpha,z}(\rho_A(\varepsilon)\Vert\sigma_A)
 \right|_{\varepsilon=0}=0.
 \label{eq:az-first-variation-zero}
\end{equation}
The explicit modular-energy--entropy realization of this stationarity is
\begin{equation}
 \Tr\!\left[
 \left(
 \overline\eta_{\alpha,z;A}(\varepsilon)
 -\sigma_A
 \right)
 K_{\sigma_A}\right]
 =\int_{I_\alpha}
 \left[
 S(\eta_{\beta,c;A}(\varepsilon))
 -S(\sigma_A)
 \right]
 \dd\mu_\alpha(\beta)
 +O(\varepsilon^2).
 \label{eq:az-first-law}
\end{equation}

The cancellation at first order may be viewed as an escort-averaged 
version of the first law of entanglement for relative entropy \cite{Blanco:2013joa}, which states that the infinitesimal change in the vacuum modular energy is equal to the infinitesimal change in the entanglement entropy to first order in perturbations around the vacuum.
The analogous statement for the $\alpha$--$z$ divergence
\eqref{eq:az-first-law} is an escort-averaged first law of entanglement:
the infinitesimal change in the escort-averaged vacuum modular energy equals
the infinitesimal change in the escort-averaged entanglement entropy to
first order around the vacuum.

Infinitesimal perturbations of the refined SRD were studied in
\cite{Bao:2019aol}, and perturbations of the SRD in free field theories in
\cite{Moosa:2020jwt,Roy:2022yzm}. The expansion of the $\alpha$--$z$ divergence was studied in \cite{May:2018tir}. Equation~\eqref{eq:az-first-law} gives
the fixed-ray $\alpha$--$z$ first law of entanglement from the integral
representation.

\subsection{Escort-averaged information metric}

At second order, one obtains
\begin{equation}
 D_{\alpha,z}(\rho_A(\varepsilon)\Vert\sigma_A)
 =\frac{\varepsilon^2}{2}
 \int_{I_\alpha}
 \Tr\!\left(
 \xi_{\beta,c;A}
 \mathcal T_{\sigma_A}(\xi_{\beta,c;A})
 \right)
 \dd\mu_\alpha(\beta)
 +O(\varepsilon^3).
 \label{eq:az-second-order-bkm-average}
\end{equation}
The integrand is the Bogoliubov--Kubo--Mori (BKM) metric evaluated on the
linearized escort perturbation.  The BKM metric is the second-order
variation of relative entropy
\cite{petz1993bogoliubov,Petz1994_CanonicalCorrelation,Lesniewski1999}.
Thus the second-order variation of the $\alpha$--$z$ divergence is an escort
average of BKM metrics evaluated on the linearized fixed-ray escorts.  The
general $\alpha$--$z$ information matrix was independently derived in
\cite{May:2018tir,Wilde:2025nki}, while its sandwiched slice was studied in
\cite{Takahashi_2017}.  The fixed-ray representation identifies the same
metric as an escort average of BKM forms and leads directly to the explicit kernel
below.

The dependence of the escort tangent on both $\alpha$ and $z$ can be made
explicit.  Put
\begin{equation}
 a_{\beta,c}\defeq\frac{1-\beta}{2c\beta},
 \qquad
 B_{\beta,c}(\varepsilon)
 \defeq
 \sigma_A^{a_{\beta,c}}
 \rho_A(\varepsilon)^{1/c}
 \sigma_A^{a_{\beta,c}}.
 \label{eq:az-perturbative-factor}
\end{equation}
Then
\begin{equation}
 B_{\beta,c}(0)=\sigma_A^{1/(c\beta)},
 \qquad
 \eta_{\beta,c;A}(\varepsilon)
 =\frac{B_{\beta,c}(\varepsilon)^{c\beta}}
 {\Tr[B_{\beta,c}(\varepsilon)^{c\beta}]}.
\end{equation}
Diagonalize the vacuum density matrix,
\begin{equation}
 \sigma_A=\sum_a p_a\lvert a\rangle\!\langle a\rvert,
 \qquad p_a>0.
 \label{eq:az-vacuum-spectrum}
\end{equation}
The derivative of the unnormalized numerator at $\varepsilon=0$ is
\begin{equation}
 D(t^{c\beta})_{\sigma_A^{1/(c\beta)}}
 \!\left[
 \sigma_A^{a_{\beta,c}}
 D(t^{1/c})_{\sigma_A}[X_A]
 \sigma_A^{a_{\beta,c}}
 \right].
 \label{eq:az-unnormalized-escort-tangent}
\end{equation}
Taking the trace of the preceding expression and using
\[
 \Tr\!\left[D(t^q)_A[Y]\right]
 =
 q\,\Tr\!\left[A^{q-1}Y\right],
\]
we obtain
\begin{align}
 &\Tr\!\left[
 D(t^{c\beta})_{\sigma_A^{1/(c\beta)}}
 \left[
   \sigma_A^{a_{\beta,c}}
   D(t^{1/c})_{\sigma_A}[X_A]
   \sigma_A^{a_{\beta,c}}
 \right]
 \right]
 \notag\\
 &\qquad
 =
 c\beta\,
 \Tr\!\left[
   \sigma_A^{\,1-1/c}
   D(t^{1/c})_{\sigma_A}[X_A]
 \right]
 \notag\\
 &\qquad
 =
 c\beta\sum_a
 p_a^{\,1-1/c}
 \left(\frac{1}{c}p_a^{\,1/c-1}\right)
 (X_A)_{aa}
 =
 \beta\,\Tr X_A
 =
 0.
 \label{eq:trace-escort-numerator-derivative}
\end{align}
In the third line, we used the eigenbasis \eqref{eq:az-vacuum-spectrum}.
Since \eqref{eq:az-unnormalized-escort-tangent} is traceless, it receives no linear correction from normalization. Therefore
\begin{equation}
 \xi_{\beta,c;A}
 =D(t^{c\beta})_{\sigma_A^{1/(c\beta)}}
 \!\left[
 \sigma_A^{a_{\beta,c}}
 D(t^{1/c})_{\sigma_A}[X_A]
 \sigma_A^{a_{\beta,c}}
 \right].
 \label{eq:az-escort-tangent-frechet}
\end{equation}

The standard divided-difference formula for Fr\'echet derivatives (see Appendix~\ref{app:frechet-derivatives}) gives
\begin{equation}
 (\xi_{\beta,c;A})_{ab}
 =m_{\beta,c}(p_a,p_b)(X_A)_{ab},
 \label{eq:az-escort-tangent-spectral}
\end{equation}
where
\begin{equation}
 m_{\beta,c}(x,y)
 =\begin{cases}
 \displaystyle
 (xy)^{\frac{1-\beta}{2c\beta}}
 \frac{x^{1/c}-y^{1/c}}
 {x^{1/(c\beta)}-y^{1/(c\beta)}},
 &x\neq y,\\[1.4ex]
 \beta,&x=y.
 \end{cases}
 \label{eq:az-escort-multiplier}
\end{equation}
For $x\neq y$, writing $r=\log(x/y)$ gives the particularly simple form
\begin{equation}
 m_{\beta,c}(x,y)
 =\frac{\sinh\!\left(\frac{r}{2c}\right)}
 {\sinh\!\left(\frac{r}{2c\beta}\right)}.
 \label{eq:az-escort-multiplier-hyperbolic}
\end{equation}
The continuous diagonal value in \eqref{eq:az-escort-multiplier} is indeed
$\beta$, and $m_{1,c}(x,y)=1$, consistently with
$\rho_1^{(c)}=\rho$.

In the eigenbasis \eqref{eq:az-vacuum-spectrum}, the BKM map has the
spectral action
\begin{equation}
 \bigl(\mathcal T_{\sigma_A}(U)\bigr)_{ab}
 =\ell(p_a,p_b)U_{ab},
 \qquad
 \ell(x,y)=
 \begin{cases}
 \displaystyle\frac{\log x-\log y}{x-y},&x\neq y,\\[1.2ex]
 \displaystyle\frac1x,&x=y.
 \end{cases}
 \label{eq:az-bkm-kernel}
\end{equation}
Using \eqref{eq:az-escort-tangent-frechet} and \eqref{eq:az-bkm-kernel}, it follows from \eqref{eq:az-second-order-bkm-average} that
\begin{equation}
 D_{\alpha,z}(\rho_A(\varepsilon)\Vert\sigma_A)
 =\frac{\varepsilon^2}{2}
 \sum_{a,b}\Gamma_{\alpha,z}(p_a,p_b)
 |(X_A)_{ab}|^2+O(\varepsilon^3),
 \label{eq:az-quadratic-spectral}
\end{equation}
where
\begin{equation}
 \Gamma_{\alpha,z}(x,y)
 \defeq
 \ell(x,y)\int_{I_\alpha}m_{\beta,c}(x,y)^2
 \dd\mu_\alpha(\beta),
 \qquad c=\frac z\alpha.
 \label{eq:az-gamma-integral}
\end{equation}

For completeness, the $\beta$ integral can be evaluated exactly.  For
$x\neq y$, put
\begin{equation}
 r=\log\frac{x}{y},
 \qquad
 A=\frac{r}{2c}=\frac{\alpha r}{2z}.
\end{equation}
Using $u=A/\beta$ gives
\begin{align}
 \int_{I_\alpha}m_{\beta,c}(x,y)^2
 \dd\mu_\alpha(\beta)
 &=\frac{\alpha}{\alpha-1}
 \int_1^\alpha
 \frac{\dd\beta}{\beta^2}
 \frac{\sinh^2 A}{\sinh^2(A/\beta)}
 \notag\\
 &=\frac{\alpha}{\alpha-1}
 \frac{\sinh^2 A}{A}
 \left[\coth\!\left(\frac A\alpha\right)-\coth A\right].
 \label{eq:az-multiplier-integral}
\end{align}
The first line is understood as an oriented integral when $\alpha<1$.
Combining \eqref{eq:az-multiplier-integral} with
$\ell(x,y)=r/[2\sqrt{xy}\sinh(r/2)]$ yields the explicit metric
\begin{equation}
 \Gamma_{\alpha,z}(x,y)
 =\frac{z}{\alpha-1}\frac1{\sqrt{xy}}
 \frac{
 \sinh\!\left(\frac{\alpha}{2z}\log\frac{x}{y}\right)
 \sinh\!\left(\frac{\alpha-1}{2z}\log\frac{x}{y}\right)}
 {
 \sinh\!\left(\frac12\log\frac{x}{y}\right)
 \sinh\!\left(\frac1{2z}\log\frac{x}{y}\right)},
 \quad x\neq y.
 \label{eq:az-gamma-closed}
\end{equation}
Its continuous diagonal value is
\begin{equation}
 \Gamma_{\alpha,z}(x,x)=\frac{\alpha}{x}.
 \label{eq:az-gamma-diagonal}
\end{equation}
The same unnormalized spectral kernel was obtained previously by May and Hijano in
\cite{May:2018tir}. In our notation, their spectral weight
is
\begin{equation}
 \Gamma^{\mathrm{MH}}_{\alpha,z}(x,y)
 :=
 \frac{z}{1-\alpha}\,
 \frac{
   \bigl(x^{\alpha/z}-y^{\alpha/z}\bigr)
   \bigl(x^{(1-\alpha)/z}-y^{(1-\alpha)/z}\bigr)
 }{
   (x-y)\bigl(x^{1/z}-y^{1/z}\bigr)
 }.
 \label{eq:may-hijano-kernel}
\end{equation}
Indeed, using
\begin{equation}
 x^{q}-y^{q}
 =
 2(xy)^{q/2}
 \sinh\!\left(\frac{q}{2}\log\frac{x}{y}\right),
\end{equation}
one obtains
\begin{equation}
 \Gamma^{\mathrm{MH}}_{\alpha,z}(x,y)
 =
 \frac{z}{\alpha-1}\frac{1}{\sqrt{xy}}\,
 \frac{
   \sinh\!\left(\frac{\alpha}{2z}\log\frac{x}{y}\right)
   \sinh\!\left(\frac{\alpha-1}{2z}\log\frac{x}{y}\right)
 }{
   \sinh\!\left(\frac{1}{2}\log\frac{x}{y}\right)
   \sinh\!\left(\frac{1}{2z}\log\frac{x}{y}\right)
 }
 =
 \Gamma_{\alpha,z}(x,y),
\end{equation}
with the same continuous diagonal value
$\Gamma^{\mathrm{MH}}_{\alpha,z}(x,x)=\alpha/x$.
Thus the kernel appearing in the Hessian agrees exactly with the
spectral kernel of \cite{May:2018tir}, while the normalized
Morozova--Chentsov kernel used below is
$\Gamma_{\alpha,z}/\alpha$.\footnote{Equation~(2.11) of
\cite{May:2018tir} defines the susceptibility as one half of the
Hessian, whereas its spectral formulas, Eqs.~(2.12) and~(A.11), identify
the displayed kernel with the full Hessian.  We compare here directly
with the spectral kernel and avoid this factor-of-two
ambiguity.}

It is standard to remove the classical R\'enyi normalization and define
\begin{equation}
 \mathcal I_{\alpha,z;\sigma_A}(X_A)
 \defeq
 \frac1\alpha\int_{I_\alpha}
 \Tr\!\left[
 \xi_{\beta,c;A}\mathcal T_{\sigma_A}(\xi_{\beta,c;A})
 \right]\dd\mu_\alpha(\beta).
 \label{eq:az-information-metric}
\end{equation}
This normalization matches that in \cite{Takahashi_2017,Wilde:2025nki}.
Then
\begin{equation}
 D_{\alpha,z}(\sigma_A+\varepsilon X_A\Vert\sigma_A)
 =\frac{\alpha\varepsilon^2}{2}
 \mathcal I_{\alpha,z;\sigma_A}(X_A)
 +O(\varepsilon^3).
 \label{eq:az-information-expansion}
\end{equation}
The normalized kernel $\Gamma_{\alpha,z}/\alpha$ has the symmetric
homogeneous form used in the Morozova--Chentsov--Petz classification
\begin{equation}
 \frac1\alpha\Gamma_{\alpha,z}(x,y)
 =\frac{1}{y f_{\alpha,z}(x/y)},
 \label{eq:az-morozova-chentsov}
\end{equation}
where
\begin{equation}
 f_{\alpha,z}(t)
 =\frac{\alpha(\alpha-1)}{z}\sqrt t\,
 \frac{
 \sinh\!\left(\frac12\log t\right)
 \sinh\!\left(\frac1{2z}\log t\right)}
 {
 \sinh\!\left(\frac{\alpha}{2z}\log t\right)
 \sinh\!\left(\frac{\alpha-1}{2z}\log t\right)}.
 \label{eq:az-representing-function}
\end{equation}
The continuous value is $f_{\alpha,z}(1)=1$, and
$f_{\alpha,z}(t)=t f_{\alpha,z}(t^{-1})$.  

Since $\dd\mu_\alpha$ is a positive measure and the BKM quadratic form is
positive semidefinite, the representation
\eqref{eq:az-information-metric} immediately implies positivity of
$\mathcal I_{\alpha,z;\sigma}$ whenever the fixed-ray construction is defined.
This pointwise positivity of the Hessian
at $\rho=\sigma$ does not require data processing. 
Data processing of the divergence implies contractivity of the information metric throughout the data-processing region. This gives a sufficient condition and does not determine the full contractivity domain. 

Let $\Phi$ be a quantum channel (a completely positive trace preserving map), and
set
\begin{equation}
 \sigma_\Phi\defeq\Phi(\sigma_A),
 \qquad
 X_\Phi\defeq\Phi(X_A),
\end{equation}
and assume that $\sigma_A$ and $\sigma_\Phi$ are faithful. Then, for sufficiently small $\varepsilon$, DPI applied to
$\rho_A(\varepsilon)=\sigma_A+\varepsilon X_A$ gives
\[
 D_{\alpha,z}
 \bigl(\sigma_\Phi+\varepsilon X_\Phi\Vert\sigma_\Phi\bigr)
 \leq
 D_{\alpha,z}
 \bigl(\sigma_A+\varepsilon X_A\Vert\sigma_A\bigr).
\]
Using \eqref{eq:az-information-expansion} on the two sides, we obtain
\[
 0\leq
 \frac{\alpha\varepsilon^2}{2}
 \left[
  \mathcal I_{\alpha,z;\sigma_A}(X_A)
  -\mathcal I_{\alpha,z;\sigma_\Phi}(X_\Phi)
 \right]
 +O(\varepsilon^3).
\]
Dividing by $\alpha\varepsilon^2/2$ and taking
$\varepsilon\to0$ proves that the escort averaged metrics are monotonic under quantum channels:
\begin{equation}
 \mathcal I_{\alpha,z;\sigma_\Phi}(X_\Phi)
 \leq
 \mathcal I_{\alpha,z;\sigma_A}(X_A).
\end{equation}
Hence
\eqref{eq:az-information-metric} is a monotone quantum Fisher metric in the
sense of Morozova--Chentsov and Petz
\cite{MorozovaChentsov1991,Petz:1999xrh,petz1996geometries}.  
Equivalently,
$f_{\alpha,z}$ is then operator monotone.  

If $[X_A,\sigma_A]=0$, choose a common eigenbasis in which $X_A$ and $\sigma_A$ are simultaneously diagonal, then
\begin{equation}
 \mathcal I_{\alpha,z;\sigma_A}(X_A)
 =\sum_a\frac{(X_A)_{aa}^2}{p_a},
 \label{eq:az-classical-fisher}
\end{equation}
independently of $z$.  Accordingly,
\begin{equation}
 D_{\alpha,z}(\sigma_A+\varepsilon X_A\Vert\sigma_A)
 =\frac{\alpha\varepsilon^2}{2}
 \sum_a\frac{(X_A)_{aa}^2}{p_a}+O(\varepsilon^3)
\end{equation}
in the commuting sector, as expected, because the classical
$\alpha$--$z$ divergence is independent of $z$.

\subsection{Vacuum balls and an escort-averaged Bekenstein bound}
\label{subsec:az-vacuum-balls}

Let $\sigma_B$ denote the vacuum state of a $d$-dimensional CFT reduced to
the ball
\begin{equation}
 B=\{\bm x:|\bm x|<R\}
\end{equation}
on the $t=0$ slice, represented in the regulated density-matrix framework
specified above.  The exact continuum modular Hamiltonian in this framework is local \cite{Casini:2011kv}
\begin{equation}
 K_{\sigma_B}
 =2\pi\int_B\dd^{d-1}x\,
 \frac{R^2-|\bm x|^2}{2R}T_{00}(\bm x),
 \label{eq:az-ball-modular-hamiltonian}
\end{equation}
up to an additive constant.
The additive constant drops out of normalized expectation-value
differences.  In what follows, the same regulator is used for all density
matrices, escort states, entropies, and modular-energy expectation values.

Let $\rho_B$ be an excited state on the ball. Equation \eqref{eq:az-modular-free-energy} becomes
\begin{align}
 D_{\alpha,z}(\rho_B\Vert\sigma_B)
 ={}&2\pi\int_B\dd^{d-1}x\,
 \frac{R^2-|\bm x|^2}{2R}
 \left(
 \langle T_{00}(\bm x)\rangle_{\overline\rho_{\alpha,z;B}}
 -\langle T_{00}(\bm x)\rangle_{\sigma_B}
 \right)
 \notag\\
 &-\left[
 \int_{I_\alpha}S((\rho_B)_\beta^{(c)})
 \dd\mu_\alpha(\beta)-S(\sigma_B)
 \right].
 \label{eq:az-ball-decomposition}
\end{align}
Here
\begin{equation}
 \overline\rho_{\alpha,z;B}
 =\int_{I_\alpha}(\rho_B)_\beta^{(c)}
 \dd\mu_\alpha(\beta)
\end{equation}
is the escort-averaged state.
Nonnegativity of each ordinary relative entropy in
\eqref{eq:az-fixed-ray-integral} gives
\begin{equation}
 \begin{aligned}
 2\pi\int_B\dd^{d-1}x\,
 \frac{R^2-|\bm x|^2}{2R}
 \left(
 \langle T_{00}(\bm x)\rangle_{\overline\rho_{\alpha,z;B}}
 -\langle T_{00}(\bm x)\rangle_{\sigma_B}
 \right)
 \geq{}&
 \int_{I_\alpha}S((\rho_B)_\beta^{(c)})
 \dd\mu_\alpha(\beta)-S(\sigma_B).
 \end{aligned}
 \label{eq:az-bekenstein-bound}
\end{equation}
This is an escort-averaged analogue of the Bekenstein-type inequalities obtained from ordinary relative entropy \cite{Casini:2008cr,Blanco:2013joa}.
Thus the escort-averaged, vacuum-subtracted ball modular-energy
is bounded below by the average vacuum-subtracted von Neumann entropy of
the escort states.

For $\alpha>1$, raywise monotonicity of the escort relative-entropy profile
\cite{Kibe:2026dtz} also gives
\begin{equation}
 D_{\alpha,z}(\rho\Vert\sigma)\geq D(\rho\Vert\sigma).
 \label{eq:az-upper-comparison}
\end{equation}
Indeed, $\beta\mapsto D(\rho_\beta^{(c)}\Vert\sigma)$ is nondecreasing \cite{Kibe:2026dtz},
and \eqref{eq:az-fixed-ray-integral} averages it over $[1,\alpha]$ while
its value at $\beta=1$ is $D(\rho\Vert\sigma)$.  Subtracting the ordinary
modular-energy--entropy decomposition of the relative entropy from \eqref{eq:az-modular-free-energy} yields
the stronger comparison for $\alpha>1$
\begin{equation}
 \Tr[(\overline\rho_{\alpha,z}-\rho)K_\sigma]
 \geq S_{\alpha,z}^{\esc}(\rho\Vert\sigma)-S(\rho).
 \label{eq:az-strong-bekenstein-general}
\end{equation}
For a vacuum ball this is
\begin{equation}
 \begin{aligned}
 2\pi\int_B\dd^{d-1}x\,
 \frac{R^2-|\bm x|^2}{2R}
 \left(
 \langle T_{00}(\bm x)\rangle_{\overline\rho_{\alpha,z;B}}
 -\langle T_{00}(\bm x)\rangle_{\rho_B}
 \right)
 \geq
 \int_{I_\alpha}S((\rho_B)_\beta^{(c)})
 \dd\mu_\alpha(\beta)-S(\rho_B).
 \end{aligned}
 \label{eq:az-strong-ball-bound}
\end{equation}
Both sides depend
on the fixed-ray escort ensemble: the energy is evaluated in its
escort-averaged state, while the entropy is averaged over its individual members.
The bounds should therefore be viewed as consequences of the
modular-energy--entropy decomposition of the escort ensemble associated with the pair
$(\rho_B,\sigma_B)$.

\subsection{Sandwiched specialization}
\label{subsec:qft-sandwiched-slice}

On the sandwiched line $z=\alpha$, one has $c=1$. The first law
\eqref{eq:az-first-law} and the vacuum-ball results of
Section~\ref{subsec:az-vacuum-balls} apply with this choice.
The linear response of an escort is given by
\eqref{eq:az-escort-tangent-spectral}, with the simplified multiplier
\begin{equation}
 m_{\beta,1}(x,y)
 =
 \begin{cases}
 \displaystyle
 (xy)^{\frac{1-\beta}{2\beta}}
 \frac{x-y}{x^{1/\beta}-y^{1/\beta}},&x\neq y,\\[1.2ex]
 \beta,&x=y.
 \end{cases}
 \label{eq:qft-sandwiched-multiplier-specialization}
\end{equation}
The corresponding kernel of the divergence Hessian is
\begin{equation}
 \Gamma_{\alpha,\alpha}(x,y)
 =\frac{\alpha}{\alpha-1}\frac1{\sqrt{xy}}
 \frac{
 \sinh\!\left(\frac{\alpha-1}{2\alpha}\log\frac{x}{y}\right)}
 {\sinh\!\left(\frac1{2\alpha}\log\frac{x}{y}\right)},
 \qquad x\neq y,
 \label{eq:qft-sandwiched-kernel-specialization}
\end{equation}
with continuous diagonal value
$\Gamma_{\alpha,\alpha}(x,x)=\alpha/x$.
The normalized information metric
$\mathcal I_{\alpha,\alpha;\sigma_A}$ has kernel
$\Gamma_{\alpha,\alpha}/\alpha$. Its representing function in
\eqref{eq:az-morozova-chentsov} takes the form
\begin{equation}
 f_{\alpha,\alpha}(t)
 =(1-\alpha)
 \frac{t^{1/\alpha}-1}{t^{(1-\alpha)/\alpha}-1},
 \label{eq:qft-sandwiched-representing-function}
\end{equation}
where $f_{\alpha,\alpha}(1)=1$ by continuity.

For $\alpha\in[1/2,1)\cup(1,\infty)$, data processing of the
sandwiched divergence implies that this metric is contractive under
quantum channels. In the limit $\alpha\to1$, the representing
function becomes $f_{1,1}(t)=(t-1)/\log t$, and the metric reduces
to the Bogoliubov--Kubo--Mori metric of ordinary relative entropy.

\subsection{Exact encoding and holographic interpretation}
\label{subsec:az-holographic-outlook}

We show that escort formation commutes with the reduced encoding
channel in a finite-dimensional operator-algebra code with exact
complementary recovery. The proof uses the standard block structure
of these codes \cite{Harlow:2016vwg,harlow2018tasi}.

Let the physical Hilbert space be
$\mathcal H_A\otimes\mathcal H_{\bar A}$. The code subspace
decomposes as
\begin{equation}
 \mathcal H_{\mathrm{code}}
 =\bigoplus_\gamma
 \mathcal H_{a_\gamma}\otimes\mathcal H_{\bar a_\gamma}.
\end{equation}
The logical algebra reconstructed from $A$ acts on the factors
$\mathcal H_{a_\gamma}$ and trivially on
$\mathcal H_{\bar a_\gamma}$. We represent a state on this algebra by
$\tau_a=\bigoplus_\gamma\tau_{a_\gamma}$, where the blocks are
positive and $\sum_\gamma\Tr\tau_{a_\gamma}=1$.
Thus $\tau_a$ describes the reduced logical state. Under exact
complementary recovery, its reduced encoding has the form
\begin{equation}
 \mathcal N_A(\tau_a)
 =U_A\left[
  \bigoplus_\gamma
  \tau_{a_\gamma}\otimes\chi_\gamma
 \right]U_A^\dagger,
 \label{eq:az-exact-code-block-form}
\end{equation}
where $U_A$ is a fixed isometry into $\mathcal H_A$, and each
$\chi_\gamma$ is a normalized state independent of $\tau_a$.
All traces on the logical algebra below are sums of the ordinary
traces over its blocks.

Let $\rho_a$ and $\sigma_a$ be two such logical states, with
$\supp\rho_a\subseteq\supp\sigma_a$. For every $\beta,c>0$, we
claim that
\begin{equation}
 \mathcal N_A\!\left((\rho_a)_\beta^{(c)}\right)
 =\bigl(\mathcal N_A(\rho_a)\bigr)_\beta^{(c)}.
 \label{eq:az-exact-code-escort-intertwining}
\end{equation}
The escort on the left uses $\sigma_a$ as its reference state.
The escort on the right uses $\mathcal N_A(\sigma_a)$. Thus the
statement compares forming an escort before encoding with forming
it from the encoded pair of states.

To prove the claim, write
\begin{equation}
 \rho_a=\bigoplus_\gamma p_\gamma\rho_{a_\gamma},
 \qquad
 \sigma_a=\bigoplus_\gamma q_\gamma\sigma_{a_\gamma},
\end{equation}
with normalized states in each nonzero block. Sectors with
$q_\gamma=0$ can be omitted by support inclusion. Inverse powers
are evaluated on the relevant support corners, including the
supports of the ancillary states $\chi_\gamma$.
Set
\begin{equation}
 X_{\beta,\gamma}^{(c)}
 \defeq
 (q_\gamma\sigma_{a_\gamma})^{a_{\beta,c}}
 (p_\gamma\rho_{a_\gamma})^{1/c}
 (q_\gamma\sigma_{a_\gamma})^{a_{\beta,c}},
 \qquad
 a_{\beta,c}=\frac{1-\beta}{2c\beta}.
\end{equation}
The same ancillary state appears in both encoded arguments. Its
powers therefore combine according to
$2a_{\beta,c}+1/c=1/(c\beta)$, giving
\begin{equation}
 \begin{aligned}
 &\mathcal N_A(\sigma_a)^{a_{\beta,c}}
  \mathcal N_A(\rho_a)^{1/c}
  \mathcal N_A(\sigma_a)^{a_{\beta,c}}
 \\
 &\qquad=
 U_A\left[
  \bigoplus_\gamma
  X_{\beta,\gamma}^{(c)}\otimes\chi_\gamma^{1/(c\beta)}
 \right]U_A^\dagger.
 \end{aligned}
 \label{eq:az-exact-code-ray-factor}
\end{equation}
Raising this operator to the power $c\beta$ restores the ancillary
state $\chi_\gamma$ in each block. Hence
\begin{equation}
 \begin{aligned}
 &\left(
  \mathcal N_A(\sigma_a)^{a_{\beta,c}}
  \mathcal N_A(\rho_a)^{1/c}
  \mathcal N_A(\sigma_a)^{a_{\beta,c}}
 \right)^{c\beta}
 \\
 &\qquad=
 U_A\left[
  \bigoplus_\gamma
  (X_{\beta,\gamma}^{(c)})^{c\beta}\otimes\chi_\gamma
 \right]U_A^\dagger
 \\
 &\qquad=
 \mathcal N_A\!\left[
  \left(
   \sigma_a^{a_{\beta,c}}\rho_a^{1/c}
   \sigma_a^{a_{\beta,c}}
  \right)^{c\beta}
 \right].
 \end{aligned}
 \label{eq:az-exact-code-unnormalized-intertwining}
\end{equation}
Trace preservation makes the two normalization factors equal.
Dividing by this common trace proves
\eqref{eq:az-exact-code-escort-intertwining}.

Linearity of the channel then gives the corresponding identity for
the average escort state,
\begin{equation}
 \mathcal N_A\!\left(
  \int_{I_\alpha}(\rho_a)_\beta^{(c)}\dd\mu_\alpha(\beta)
 \right)
 =\int_{I_\alpha}
  \bigl(\mathcal N_A(\rho_a)\bigr)_\beta^{(c)}
  \dd\mu_\alpha(\beta).
 \label{eq:az-exact-code-barycenter-intertwining}
\end{equation}
Relative entropy is also preserved. Its trace formula is a sum
over the subnormalized blocks, including their probability weights.
Adjoining the same normalized ancillary state to both arguments
in a block leaves its contribution unchanged. Thus each escort obeys
\begin{equation}
 D\!\left(
  \bigl(\mathcal N_A(\rho_a)\bigr)_\beta^{(c)}
  \middle\Vert\mathcal N_A(\sigma_a)
 \right)
 =D\!\left((\rho_a)_\beta^{(c)}\Vert\sigma_a\right).
 \label{eq:az-exact-code-integrand-equality}
\end{equation}
This is the exact code-model counterpart of the bulk--boundary
relative-entropy relation \cite{Jafferis_2016}.
Integrating over $\beta$ gives equality of the logical and encoded
$\alpha$--$z$ divergences. On the sandwiched line, this recovers the
divergence equality discussed in \cite{Caginalp:2022uzd}.
Equation~\eqref{eq:az-exact-code-escort-intertwining} establishes
the corresponding relation for each individual escort state.

The information metrics agree as well. Taking
$\rho_a(\varepsilon)=\sigma_a+\varepsilon X_a$ in a fixed support
corner, with $X_a$ Hermitian and traceless, and expanding the
divergence equality at $\varepsilon=0$ gives
\begin{equation}
 \mathcal I_{\alpha,z;\mathcal N_A(\sigma_a)}
 \bigl(\mathcal N_A(X_a)\bigr)
 =\mathcal I_{\alpha,z;\sigma_a}(X_a).
\end{equation}
These statements follow from the exact encoding structure and
require no semiclassical interpretation of the escort states.

\subsubsection{Holographic outlook}
\label{subsubsec:az-holographic-interpretation}

For vacuum perturbations in a holographic CFT, the ordinary
entanglement first law for all balls, together with the holographic
entropy formula \cite{Ryu:2006bv,Hubeny:2007xt}, gives the linearized
Einstein equations about AdS at leading semiclassical order
\cite{Lashkari:2013koa,Faulkner:2013ica}. At quadratic order, the
BKM information metric is represented by bulk canonical energy
in the corresponding AdS--Rindler wedge \cite{Lashkari:2015hha}.
The escort first law averages ordinary first laws and therefore
does not by itself imply an additional gravitational equation.

Suppose that every escort tangent $\xi_{\beta,c;A}$ has a compatible
bulk representative $\delta\Phi_{\beta,c;A}$ in one common
semiclassical code subspace. Applying the canonical-energy
identification to each tangent would give, schematically,
\begin{equation}
 \alpha\,\mathcal I_{\alpha,z;\sigma_A}(X_A)
 \quad\longleftrightarrow\quad
 \int_{I_\alpha}
 \mathcal E_A^{\mathrm{can}}
 \bigl(\delta\Phi_{\beta,c;A},\delta\Phi_{\beta,c;A}\bigr)
 \dd\mu_\alpha(\beta),
 \label{eq:az-escort-averaged-canonical-energy}
\end{equation}
with canonical energy normalized to correspond to the BKM Hessian.
The factor of $\alpha$ follows from the normalization of
\eqref{eq:az-information-metric}.

For suitable states on boundary balls, replica constructions give
bulk descriptions of sandwiched escorts at integer escort orders
\cite{Bao:2019aol}.
Related calculations concern R\'enyi divergences in JT gravity
\cite{Ugajin:2020dyd} and the second-order $\alpha$--$z$ divergence
for integer $\alpha$ and arbitrary $z>0$, assuming equality of
bulk and boundary modular flows \cite{May:2018tir}.
The integral in \eqref{eq:az-escort-averaged-canonical-energy},
however, requires compatible representatives for every
$\beta\in I_\alpha$, including noninteger values even on the
sandwiched line. The exact code result above does not establish
this common semiclassical realization. The average escort state
need not describe a single geometry, and the interchange of
bulk reconstruction, regulator removal, and escort averaging
requires further control.

\section{Null shape variations and the
\texorpdfstring{$\alpha$--$z$}{alpha-z} R\'enyi QNEC}
\label{sec:az-null-variations}

In this section, we apply the modular-energy--entropy decomposition to null shape
variations. 
We first derive the complete bilocal Hessian of
$D_{\alpha,z}$ under our assumed regularization of QFT allowing the use of density matrices.  We then formulate a diagonal $\alpha$--$z$
QNEC for $\alpha>1$ in the upper data-processing region and determine its
quasi-local content.  On the sandwiched slice this reduces to the R\'enyi QNEC (RQNEC).  No positivity away from the sandwiched slice is assumed until
Conjecture~\ref{hyp:az-qnec} below. 
Under appropriate regularity assumptions, null differentiating the fixed-ray identity gives an exact escort average of relative entropy Hessians. However, the integrands are generally not ordinary QNEC variations. 
The first argument in the relative entropy integrand need not be the restriction of a fixed global state, and we explicitly identify the additional modular-energy response caused by this cut
dependence, which we call escort transport.
This implies that the R\'enyi QNEC and the conjectured {\az} QNEC are not merely escort averages of the ordinary QNEC.

Consider a local Poincar\'e-invariant QFT in $d$-dimensional Minkowski
space, with null coordinates $x^\pm$ and metric
\begin{equation}
  \dd s^2=-\dd x^+\dd x^-+\dd y^2,
\end{equation}
where $y\in\mathbb R^{d-2}$ denotes the transverse coordinates.  A cut
of the null hypersurface $x^-=0$ is specified by
\begin{equation}
  x^+=V(y).
  \label{eq:az-null-cut}
\end{equation}
Let $R[V]$ be the causal completion of the portion $x^+>V(y)$, and let $\rho_V$ and $\sigma_V$ be the states obtained by restricting
a fixed global state and the Minkowski vacuum to $R[V]$, represented in the
regulated density-matrix framework described in
Section~\ref{sec:qft}.  We assume smooth identifications of the regulated regional state spaces for nearby cuts and the existence of the first and second shape derivatives used below. One can think of the relevant density matrices as being faithful on the resulting common regulated state space. More generally, we also allow families with non-full but constant support\footnote{After identifying the nearby regional state spaces with a fixed regulated Hilbert space $\mathcal H_{\mathrm{reg}}$, cf.~\cite{Hollands:2025glm} and \cite[Lemma~3.4]{Kibe:2026wsg}, constant support means that there is a projection $P$, independent of the cut $V$, such that the relevant density matrices have support $P$ throughout the neighborhood under consideration. They are then faithful as states on the corner $P\mathcal B(\mathcal H_{\mathrm{reg}})P$, although they need not be faithful on the full ambient algebra.}, in which case all operator functions and Fr\'echet derivatives are evaluated in the fixed support corner. We restrict to a sufficiently
small neighborhood of the cut in which the positive operators entering
the functional calculus remain uniformly invertible on their
supports.\footnote{Equivalently, at each fixed regulator there is a
constant $\lambda_*>0$ such that the relevant nonzero eigenvalues are
bounded below by $\lambda_*$ for all cuts in the neighborhood.  The bound
may depend on the regulator and need not be uniform as the cutoff is removed.}  We furthermore assume the regulator and subtraction prescription to be compatible with the standard continuum null-plane modular flow, so that the centered vacuum modular generator is represented by the local stress-tensor expression below.

Use the centered vacuum modular Hamiltonian
\begin{equation}
 H[V]\defeq-\log\sigma_V-S(\sigma_V)\one,
 \qquad
 \Tr(\sigma_VH[V])=0.
 \label{eq:az-centered-null-hamiltonian}
\end{equation}
On the null plane it has the local form \cite{Casini:2017roe}
\begin{equation}
 H[V]
 =2\pi\int\dd^{d-2}y\int_{V(y)}^\infty\dd u\,
 (u-V(y))T_{++}(u,0,y).
 \label{eq:az-null-hamiltonian}
\end{equation}
A review of the origin of this formula can be found in Section 3.1 of \cite{Faulkner:2024gst}.
Define the half-ray averaged null-energy operator
\begin{equation}
 P_+[V;y]\defeq
 \int_{V(y)}^\infty\dd u\,T_{++}(u,0,y).
 \label{eq:az-half-ray-energy}
\end{equation}
Then
\begin{align}
 H_{;y}[V]
 &\defeq\frac{\delta H[V]}{\delta V(y)}
 =-2\pi P_+[V;y],
 \label{eq:az-H-first-shape}
 \\
 H_{;yy'}[V]
 &\defeq\frac{\delta^2H[V]}
 {\delta V(y)\delta V(y')}
 =2\pi T_{++}(V(y),0,y)\delta_\perp(y-y').
 \label{eq:az-H-second-shape}
\end{align}

For a scalar shape functional $\mathcal F[V]$, write
\begin{equation}
 \frac{\delta^2\mathcal F[V]}
 {\delta V(y)\delta V(y')}
 =\mathcal F''_{++}[V;y]\delta_\perp(y-y')
 +\mathcal F^{(2)}_{\off}[V;y,y'].
 \label{eq:az-contact-decomposition}
\end{equation}
The first term is the diagonal contact term and the second contains no term
supported on $y=y'$.  For simplicity, we assume a regulator and subtraction prescription in
which no transverse derivatives of $\delta_\perp$ occur; otherwise
$\mathcal F''_{++}$ denotes specifically the coefficient of the
undifferentiated delta distribution.

\subsection{Escort states and transport}
\label{subsec:az-shape-functional}

Fix $\alpha>1$, $z>0$, and $c=z/\alpha$, and assume that
$Q_{\alpha,z}(\rho_V\Vert\sigma_V)$ is finite in a neighborhood of
the cut. The upper data-processing region is
\begin{equation}
 \max\{\alpha/2,\alpha-1\}\leq z\leq\alpha.
 \label{eq:az-upper-dpi-region}
\end{equation}
This will be the parameter range of the $\alpha$--$z$ QNEC conjecture. The identities
derived below require only the fixed-ray integral representation and
the regularity assumptions stated above.

For each cut $V$ and each $\beta\in[1,\alpha]$, we construct the escort
\begin{equation}
 \eta_{\beta,c}[V]
 \defeq
 \frac{
 \left(
 \sigma_V^{\frac{1-\beta}{2c\beta}}
 \rho_V^{1/c}
 \sigma_V^{\frac{1-\beta}{2c\beta}}
 \right)^{c\beta}}
 {\Tr\!\left[\left(
 \sigma_V^{\frac{1-\beta}{2c\beta}}
 \rho_V^{1/c}
 \sigma_V^{\frac{1-\beta}{2c\beta}}
 \right)^{c\beta}\right]}.
 \label{eq:az-moving-escort}
\end{equation}
At $\beta=1$, this is the original reduced state $\rho_V$. Using the
probability measure
$\dd\mu_\alpha(\beta)=\frac{\alpha}{\alpha-1}\frac{\dd\beta}{\beta^2}$,
we define the average escort state and the average escort entropy by
\begin{align}
 \overline\eta_{\alpha,z}[V]
 &\defeq
 \int_1^\alpha\eta_{\beta,c}[V]\dd\mu_\alpha(\beta),
 \label{eq:az-moving-barycenter}
 \\
 \mathcal S_{\alpha,z}^{\esc}[V]
 &\defeq
 \int_1^\alpha S(\eta_{\beta,c}[V])\dd\mu_\alpha(\beta).
 \label{eq:az-moving-escort-entropy}
\end{align}
The second quantity averages the von Neumann entropies of the individual
escorts and is generally different from
$S(\overline\eta_{\alpha,z}[V])$. The modular-energy--entropy
decomposition \eqref{eq:az-modular-free-energy} now reads
\begin{equation}
 \mathcal D_{\alpha,z}[V]
 \defeq D_{\alpha,z}(\rho_V\Vert\sigma_V)
 =\Tr(\overline\eta_{\alpha,z}[V]H[V])
 -\mathcal S_{\alpha,z}^{\esc}[V]
 +S(\sigma_V).
 \label{eq:az-null-free-energy}
\end{equation}

Before differentiating this identity, we should explain how an escort
changes when the cut moves. Let $W(y)\geq V(y)$, so that
$R[W]\subset R[V]$, and denote restriction to the smaller region by
$\mathcal R_{V\to W}$. There are two ways to obtain an auxiliary state
on $R[W]$. We can form the escort at $V$ and then restrict it, or we can
first restrict both $\rho_V$ and $\sigma_V$ and then form the escort
at $W$. The resulting states are
\begin{equation}
 \mathcal R_{V\to W}\bigl(\eta_{\beta,c}[V]\bigr)
 \qquad\text{and}\qquad
 \eta_{\beta,c}[W].
 \label{eq:az-escort-restriction-comparison}
\end{equation}
These agree at $\beta=1$, but need not agree for $\beta>1$, since  restriction
does not generally commute with the operator powers and normalization
in \eqref{eq:az-moving-escort}. Thus, although $\rho_V$ and $\sigma_V$
are restrictions of fixed global states, the escorts constructed from
them need not have this property.

To see the effect on modular energy, first consider a family obtained
by restricting a fixed global state. If the regulated modular-energy
expectation agrees with the local formula
\eqref{eq:az-null-hamiltonian}, the stress-tensor expectation value in
that integral is independent of the cut. Differentiating the lower
limit and the weight $(u-V(y))$ twice gives $2\pi$ times the null energy
density at the cut. For the escorts, the stress-tensor expectation
value can itself change as the cut moves, because a new escort is
constructed at each cut. Its response contributes to the second
variation of the modular energy.

To identify this contribution, write
\begin{equation}
 \mathcal M_{\alpha,z}[V]
 \defeq\Tr(\overline\eta_{\alpha,z}[V]H[V]).
 \label{eq:az-escort-modular-energy}
\end{equation}
We call $\mathfrak X_{\alpha,z;++}$ in
\begin{equation}
 (\mathcal M_{\alpha,z})''_{++}[V;y]
 =2\pi\langle T_{++}(V(y),0,y)\rangle_{\overline\eta_{\alpha,z}[V]}
 +\mathfrak X_{\alpha,z;++}[V;y]
 \label{eq:az-transport-modular-response}
\end{equation}
the escort transport term. The local energy term comes from the explicit
second variation of $H[V]$. Transport contains the second response of
the escort state and the mixed response in which both the state and
$H[V]$ vary. No sign is assumed for this remainder. If the escorts are
restrictions of fixed global states and their regulated modular-energy
expectations respect the local stress-tensor formula, the remainder
vanishes. The precise compatibility condition is given in
\eqref{eq:az-strong-restriction-compatibility} below.

Equations \eqref{eq:az-null-free-energy} and
\eqref{eq:az-transport-modular-response} already explain the form of
the null-shape identity. The divergence Hessian contains the null
energy of the average escort state, the second variation of the
vacuum-subtracted average escort entropy, and transport. Nonnegativity of the diagonal contact coefficient therefore
constrains the corresponding energy, entropy, and transport
contributions.

We now introduce the responses needed to compute these terms. With
$\beta$ and $c$ held fixed, let
\begin{equation}
 \eta_{\beta,c;y}
 \defeq\frac{\delta\eta_{\beta,c}[V]}{\delta V(y)},
 \qquad
 \eta_{\beta,c;yy'}
 \defeq\frac{\delta^2\eta_{\beta,c}[V]}
 {\delta V(y)\delta V(y')}.
 \label{eq:az-escort-responses}
\end{equation}
We suppress the argument $V$ when no confusion is possible.
Differentiation under the $\beta$ integral is included in our
regularity assumptions. Explicit formulas for these derivatives
are given in Appendix~\ref{ssec:escort-response-explicit}.
Normalization implies
\begin{equation}
 \Tr\eta_{\beta,c;y}=0,
 \qquad
 \Tr\eta_{\beta,c;yy'}=0.
 \label{eq:az-moving-state-traceless}
\end{equation}

The mixed modular-energy response involves the first variation of
$H[V]$, which is $-2\pi P_+[V;y]$. It is therefore convenient to set
\begin{equation}
 \mathcal C_{\beta,c}(y,y')
 \defeq
 \Tr\!\left[
 \eta_{\beta,c;y}P_+[V;y']
 +\eta_{\beta,c;y'}P_+[V;y]
 \right].
 \label{eq:az-moving-transport-kernel}
\end{equation}
The mixed response is $-2\pi\mathcal C_{\beta,c}(y,y')$.
Together with $\Tr(\eta_{\beta,c;yy'}H)$, it accounts for the
terms in the modular-energy Hessian beyond the explicit second
variation of $H$. Averaging over $\beta$ and taking the diagonal
contact coefficient will give the transport term in
\eqref{eq:az-local-transport}.

For the entropy response, we use the logarithmic Fr\'echet derivative
$\mathcal T_\eta=D(\log)_\eta$ defined in
\eqref{eq:az-log-frechet}. Its resolvent representation and cyclicity
of the trace give
\begin{equation}
 \Tr[\eta\mathcal T_\eta(U)]
 =\Tr\!\left[U\int_0^\infty
 \eta(\eta+r\one)^{-2}\dd r\right]
 =\Tr U,
\end{equation}
which follows since the  integral is the identity on the support of $\eta$. For a variation
of a normalized state, $\Tr U=0$. Differentiating
$S(\eta)=-\Tr(\eta\log\eta)$ once and then twice consequently gives
\begin{align}
 S(\eta)_{;y}
 &=-\Tr(\eta_{;y}\log\eta),
 \label{eq:az-entropy-first-shape}
 \\
 S(\eta)_{;yy'}
 &=-\Tr(\eta_{;yy'}\log\eta)
 -\Tr\!\left[\eta_{;y}\mathcal T_\eta(\eta_{;y'})\right].
 \label{eq:az-entropy-second-shape}
\end{align}
All operator functions are evaluated in the fixed support corner
when necessary. We denote the bilinear term by the
Bogoliubov--Kubo--Mori kernel
\begin{equation}
 \mathcal B_{\beta,c}(y,y')
 \defeq
 \Tr\!\left[
 \eta_{\beta,c;y}
 \mathcal T_{\eta_{\beta,c}}(\eta_{\beta,c;y'})
 \right].
 \label{eq:az-moving-bkm-kernel}
\end{equation}
The Hessian of the average escort entropy is then
\begin{equation}
 (\mathcal S_{\alpha,z}^{\esc})_{;yy'}
 =-\int_1^\alpha\dd\mu_\alpha(\beta)
 \left\{
 \Tr[\eta_{\beta,c;yy'}\log\eta_{\beta,c}]
 +\mathcal B_{\beta,c}(y,y')
 \right\}.
 \label{eq:az-escort-entropy-hessian}
\end{equation}

\subsection{Full bilocal shape Hessian}
\label{subsec:az-full-hessian}

We now evaluate the modular-energy and entropy responses introduced
in Section~\ref{subsec:az-shape-functional}. Differentiating
\eqref{eq:az-null-free-energy} once gives
\begin{equation}
 \mathcal D_{\alpha,z;y}
 =-2\pi\langle P_+[V;y]\rangle_{\overline\eta_{\alpha,z}}
 +\int_1^\alpha\dd\mu_\alpha(\beta)
 \Tr\!\left[
 \eta_{\beta,c;y}(H+\log\eta_{\beta,c})
 \right]
 +S(\sigma)_{;y},
 \label{eq:az-first-shape-derivative}
\end{equation}
where the first term comes from the explicit variation of the modular
Hamiltonian, the integral accounts for the response of the escorts,
and the last term comes from the vacuum entropy subtraction.
Differentiating once more gives
\begin{equation}
 \begin{aligned}
 \mathcal D_{\alpha,z;yy'}
 ={}&2\pi
 \langle T_{++}(V(y),0,y)\rangle_{\overline\eta_{\alpha,z}}
 \delta_\perp(y-y')
 +S(\sigma)_{;yy'}
 \\
 &+\int_1^\alpha\dd\mu_\alpha(\beta)
 \Bigl\{
 \Tr[\eta_{\beta,c;yy'}(H+\log\eta_{\beta,c})]
 -2\pi\mathcal C_{\beta,c}(y,y')
 +\mathcal B_{\beta,c}(y,y')
 \Bigr\}.
 \end{aligned}
 \label{eq:az-full-shape-hessian}
\end{equation}
Under the regulator and differentiability assumptions stated above,
this is an exact identity.
The terms involving $\log\eta_{\beta,c}$ and $\mathcal B_{\beta,c}$
give minus the Hessian of the average escort entropy. The combination
$\Tr(\eta_{\beta,c;yy'}H)-2\pi\mathcal C_{\beta,c}(y,y')$
is the modular-energy response beyond the explicit local
stress-tensor term.

\subsubsection{Local diagonal variation}
\label{subsec:az-local-variation}

The local relation follows by equating the coefficients of
$\delta_\perp(y-y')$. Using the contact decomposition specified in
\eqref{eq:az-contact-decomposition}, we write
\begin{align}
 \eta_{\beta,c;yy'}
 &=\eta^{(2)}_{\beta,c;++}(y)\delta_\perp(y-y')
 +\eta^{(2)}_{\beta,c;\off}(y,y'),
 \notag\\
 \mathcal B_{\beta,c}(y,y')
 &=\mathcal B_{\beta,c;++}(y)\delta_\perp(y-y')
 +\mathcal B_{\beta,c;\off}(y,y'),
 \label{eq:az-response-contact-decompositions}\\
 \mathcal C_{\beta,c}(y,y')
 &=\mathcal C_{\beta,c;++}(y)\delta_\perp(y-y')
 +\mathcal C_{\beta,c;\off}(y,y').
 \notag
\end{align}
The entropy contribution is therefore
\begin{equation}
 (\mathcal S_{\alpha,z}^{\esc})''_{++}(y)
 =-\int_1^\alpha\dd\mu_\alpha(\beta)
 \left\{
 \Tr[\eta^{(2)}_{\beta,c;++}(y)\log\eta_{\beta,c}]
 +\mathcal B_{\beta,c;++}(y)
 \right\}.
 \label{eq:az-local-escort-entropy}
\end{equation}
The transport term introduced in
\eqref{eq:az-transport-modular-response} has the explicit form
\begin{equation}
 \mathfrak X_{\alpha,z;++}(y)
 =\int_1^\alpha\dd\mu_\alpha(\beta)
 \left\{
 \Tr[\eta^{(2)}_{\beta,c;++}(y)H]
 -2\pi\mathcal C_{\beta,c;++}(y)
 \right\}.
 \label{eq:az-local-transport}
\end{equation}
Combining these expressions with the local stress-tensor term in
\eqref{eq:az-full-shape-hessian} gives
\begin{equation}
 (D_{\alpha,z})''_{++}[V;y]
 =2\pi\langle T_{++}(V(y),0,y)\rangle_{\overline\eta_{\alpha,z}}
 -(\mathcal S_{\alpha,z}^{\esc})''_{++}[V;y]
 +S''_{\sigma,++}[V;y]
 +\mathfrak X_{\alpha,z;++}[V;y].
 \label{eq:az-local-free-energy}
\end{equation}
Here, the local energy is evaluated in the average escort state
$\overline\eta_{\alpha,z}[V]$, which generally has a different
stress-tensor expectation value from the original excited state
$\rho_V$. The entropy terms give minus the second variation of the
vacuum-subtracted average escort entropy.

All terms in \eqref{eq:az-local-free-energy} are evaluated with the
same regulator and subtraction prescription. A finite continuum
limit of the divergence Hessian does not by itself imply separate
finite limits for the local-energy and transport terms. When these
terms are individually singular, the identity must be interpreted
through the limit of their regulated combination.

\subsubsection{Off-diagonal variation}
\label{subsec:az-off-diagonal}

For $y\neq y'$, the full Hessian gives
\begin{equation}
 \begin{aligned}
 D^{(2)}_{\alpha,z;\off}[V;y,y']
 ={}&S^{(2)}_{\sigma;\off}[V;y,y']
 \\
 &+\int_1^\alpha\dd\mu_\alpha(\beta)
 \Bigl\{
 \Tr[\eta^{(2)}_{\beta,c;\off}(y,y')
 (H+\log\eta_{\beta,c})]
 \\
 &\hspace{31mm}
 -2\pi\mathcal C_{\beta,c;\off}(y,y')
 +\mathcal B_{\beta,c;\off}(y,y')
 \Bigr\}.
 \end{aligned}
 \label{eq:az-off-diagonal-hessian}
\end{equation}
For null-plane regions, the continuum vacuum is a Markov state, i.e., it saturates strong
subadditivity \cite{Casini:2017roe}. Consequently, mixed vacuum
entropy variations generated by deformations with disjoint transverse
support vanish, and
$S^{(2)}_{\sigma;\off}[V;y,y']=0$ for $y\neq y'$.
We retain the vacuum term in
\eqref{eq:az-off-diagonal-hessian} because a finite regulator need
not preserve the exact Markov property.

There is no explicit local stress-tensor term in this equation,
since $H_{;yy'}$ is supported on the transverse diagonal.
The remaining terms describe the response to deformations at
distinct transverse points. The diagonal QNEC conjecture
in Section~\ref{subsec:az-qnec} imposes no sign condition on this
off-diagonal response. Under its finiteness and non-rigidity
assumptions, the no-go theorem of \cite{Kibe:2026bcn} rules out
a universal off-diagonal QNEC for the Petz, sandwiched, and
$\alpha$--$z$ families in their data-processing ranges.

\subsection{Conjectural \texorpdfstring{$\alpha$--$z$}{alpha-z} R\'enyi QNEC and
its quasi-local form}
\label{subsec:az-qnec}

We now state the additional positivity condition that turns
\eqref{eq:az-local-free-energy} into an energy bound. This is a
conjecture about the continuum QFT. The regulated identities derived
above provide its energy--entropy interpretation when the relevant
limits exist.

\begin{conjecture}[Diagonal $\alpha$--$z$ R\'enyi QNEC]
\label{hyp:az-qnec}
Fix $\alpha>1$ and $z$ in the upper data-processing region
\eqref{eq:az-upper-dpi-region}. Let $\rho_V$ and $\sigma_V$ be the
restrictions of a fixed excited state and the vacuum to $R[V]$.
Suppose that $D_{\alpha,z}(\rho_V\Vert\sigma_V)$ is finite in a
neighborhood of the cut and that its diagonal second null shape
variation exists. Then
\begin{equation}
 (D_{\alpha,z})''_{++}[V;y]\geq0.
 \label{eq:az-qnec-hypothesis}
\end{equation}
\end{conjecture}

Suppose that the regulator limit of \eqref{eq:az-local-free-energy}
computes the continuum shape variation, with finite limits for the
local energy, transport, and the vacuum-subtracted entropy variation.
The conjecture is then equivalent to
\begin{equation}
 2\pi\langle T_{++}(V(y),0,y)\rangle_{\overline\eta_{\alpha,z}}
 \geq
 (\mathcal S_{\alpha,z}^{\esc})''_{++}[V;y]
 -S''_{\sigma,++}[V;y]
 -\mathfrak X_{\alpha,z;++}[V;y].
 \label{eq:az-qnec-quasilocal}
\end{equation}
This bounds the null energy of the average escort state in terms of
the second variation of the average escort entropy and the transport
contribution. In a null-plane prescription with vanishing vacuum
entropy contact term, one can set $S''_{\sigma,++}=0$. We retain it
to keep the subtraction explicit.
If the local-energy and transport terms are separately singular,
the conjecture instead constrains the limit of their combination
with the entropy terms in \eqref{eq:az-local-free-energy}.

There is a special case in which the bound reduces to an average of
ordinary QNEC inequalities. Suppose that, for each $\beta$, the
escorts at all nearby cuts are restrictions of a fixed global state
$\eta_{\beta,c}^{\mathrm{glob}}$. Suppose also that the regulated
modular-energy expectation respects the local stress-tensor formula,
so that
\begin{equation}
 \Tr(\eta_{\beta,c}[V]H[V])
 =2\pi\int\dd^{d-2}y\int_{V(y)}^\infty\dd u\,
 (u-V(y))
 \langle T_{++}(u,0,y)\rangle_{\eta_{\beta,c}^{\mathrm{glob}}}
 \label{eq:az-strong-restriction-compatibility}
\end{equation}
holds throughout this neighborhood. The stress-tensor expectation
value on the right is independent of the cut. With sufficient
regularity to differentiate twice and interchange the derivatives
with the escort average, the diagonal second variation gives only
the local energy term. Hence $\mathfrak X_{\alpha,z;++}=0$.

Whenever the ordinary QNEC applies to these global escort states in
the same prescription, averaging it over $\beta$ gives
\begin{equation}
 2\pi\langle T_{++}\rangle_{\overline\eta_{\alpha,z}}
 \geq
 (\mathcal S_{\alpha,z}^{\esc})''_{++}-S''_{\sigma,++}.
 \label{eq:az-formal-average-qnec}
\end{equation}
The escort construction does not generally satisfy these restriction
conditions. The transport term in \eqref{eq:az-qnec-quasilocal}
accounts for the additional modular-energy response in that case.

One can also express the bound using the entropy of the average
escort state. The difference between this entropy and the average
of the individual entropies is the Holevo information
\begin{equation}
 \chi_{\alpha,z}^{\esc}[V]
 \defeq
 S(\overline\eta_{\alpha,z}[V])
 -\mathcal S_{\alpha,z}^{\esc}[V].
 \label{eq:az-moving-holevo}
\end{equation}
Substitution in \eqref{eq:az-qnec-quasilocal} gives
\begin{equation}
 2\pi\langle T_{++}\rangle_{\overline\eta_{\alpha,z}}
 \geq
 S(\overline\eta_{\alpha,z})''_{++}
 -(\chi_{\alpha,z}^{\esc})''_{++}
 -S''_{\sigma,++}
 -\mathfrak X_{\alpha,z;++}.
 \label{eq:az-qnec-holevo}
\end{equation}
The Holevo term records the change from averaging entropies to taking
the entropy after averaging. Although
$\chi_{\alpha,z}^{\esc}\geq0$ at fixed regulator, its second shape
variation has no sign fixed by this inequality. The transport term
also remains present, since the average escort state need not be the
restriction of a fixed global state. Both terms distinguish
\eqref{eq:az-qnec-holevo} from an ordinary QNEC for the average escort state.

\subsubsection{Sandwiched specialization}
\label{subsec:null-sandwiched-slice}

On the sandwiched line $z=\alpha$, one has $c=1$. We write
$D_\alpha=D_{\alpha,\alpha}$ and use a superscript $\mathrm{s}$
for the escort quantities evaluated at $c=1$, with the redundant
parameter suppressed. The diagonal identity
\eqref{eq:az-local-free-energy} becomes
\begin{equation}
 (D_\alpha)''_{++}[V;y]
 =2\pi
 \langle T_{++}(V(y),0,y)\rangle_{\overline\eta_\alpha^{\,\mathrm{s}}}
 -(\mathcal S_\alpha^{\esc,\mathrm{s}})''_{++}[V;y]
 +S''_{\sigma,++}[V;y]
 +\mathfrak X_{\alpha,++}^{\,\mathrm{s}}[V;y].
 \label{eq:null-sandwiched-local-identity}
\end{equation}
The transport term is the specialization
\begin{equation}
 \mathfrak X_{\alpha,++}^{\,\mathrm{s}}[V;y]
 \defeq\mathfrak X_{\alpha,\alpha;++}[V;y],
 \label{eq:null-sandwiched-transport}
\end{equation}
and its integral expression is \eqref{eq:az-local-transport} at $c=1$.

The diagonal R\'enyi QNEC is the positivity statement
\begin{equation}
 (D_\alpha)''_{++}[V;y]\geq0.
 \label{eq:null-sandwiched-rqnec}
\end{equation}
Whenever \eqref{eq:null-sandwiched-rqnec} holds, the diagonal identity
gives the energy bound \eqref{eq:az-qnec-quasilocal} at $z=\alpha$,
with the continuum qualifications stated in
Section~\ref{subsec:az-qnec}. The full bilocal identity and the Holevo
form are likewise obtained from \eqref{eq:az-full-shape-hessian}
and \eqref{eq:az-qnec-holevo}.

\section{Coherent-state escorts and the \texorpdfstring{$\alpha$--$z$}{az} R\'enyi QNEC for a free scalar}
\label{sec:null-coherent-transport}

We now compute the $\alpha$--$z$ divergence  of a coherent free-scalar excitation with respect to the vacuum, and construct the relevant escorts that appear in the integral representation.

Consider a canonically normalized, free massive real scalar field in
$d>2$ spacetime dimensions. We use light-cone coordinates
\begin{equation}
 x^\pm=x^0\pm x^1,
 \qquad
 \dd s^2=-\dd x^+\dd x^-+\dd y^2,
 \qquad
 y\in\mathbb R^{d-2},
 \label{eq:null-fibre-coordinates}
\end{equation}
and choose the null plane $x^-=0$. Its generators are labelled by $y$,
and $u\defeq x^+$ is an affine coordinate along each generator. With
$p^\pm=p^0\pm p^1$, the mass shell and invariant measure are
\begin{equation}
 p^+p^-=m^2+|p_\perp|^2,
 \qquad
 \frac{\dd^{d-1}p}{(2\pi)^{d-1}2p^0}
 =\frac{\dd p^-}{2p^-}
  \frac{\dd^{d-2}p_\perp}{(2\pi)^{d-1}},
 \qquad p^->0.
 \label{eq:null-fibre-measure}
\end{equation}
The mass enters through $p^+=(m^2+|p_\perp|^2)/p^-$, which is conjugate
to $x^-$. On the null plane, the plane-wave phase is therefore
\begin{equation}
 p\cdot\mathbf x\big|_{x^-=0}
 =-\frac12p^-u+p_\perp\cdot y,
 \label{eq:null-fibre-phase}
\end{equation}
with no dependence on $m$.

Formally restricting the vacuum two-point function to this plane and
performing the transverse momentum integral gives
\begin{equation}
 \left\langle\phi(u,0,y)\phi(u',0,y')\right\rangle_\Omega
 =\frac{\delta^{d-2}(y-y')}{4\pi}
  \int_0^\infty\frac{\dd p^-}{p^-}\,
  e^{-\frac{i}{2}p^-(u-u'-i0)}.
 \label{eq:null-fibre-formal-scalar-two-point}
\end{equation}
We say ``formally''  because the integral diverges at zero
longitudinal momentum. For a smooth, compactly supported longitudinal
smearing $h$, the vacuum variance contains the integral
\begin{equation}
 \int_0^\infty\frac{\dd k}{k}\,|\widehat h(k)|^2,
 \qquad
 \widehat h(k)\defeq\int_{\mathbb R}\dd u\,e^{-iku}h(u),
 \label{eq:null-fibre-scalar-smearing-norm}
\end{equation}
where $k=p^-/2$. This is finite only if
$\widehat h(0)=\int h(u)\,\dd u=0$. For such a test function,
$\widehat h(k)=O(k)$ near the origin, which removes the infrared
divergence. Thus the scalar can be restricted to the null plane on
test functions with vanishing longitudinal zero mode. Its derivative
$\partial_u\phi$ has no such obstruction, since differentiation
removes the constant mode.

We define the current with the normalization
\begin{equation}
 J(u,y)\defeq\sqrt{2}\,\partial_u\phi(u,0,y).
 \label{eq:null-fibre-current-definition}
\end{equation}
Differentiating the two-point function gives
\begin{equation}
 \left\langle J(u,y)J(u',y')\right\rangle_\Omega
 =-\frac{1}{2\pi}
  \frac{\delta^{d-2}(y-y')}{(u-u'-i0)^2}.
 \label{eq:null-fibre-current-two-point}
\end{equation}
The difference of the two boundary values in
\eqref{eq:null-fibre-formal-scalar-two-point} gives the commutator
\begin{equation}
 [\phi(u,0,y),\phi(u',0,y')]
 =-\frac{i}{4}\operatorname{sgn}(u-u')\,
  \delta^{d-2}(y-y').
 \label{eq:null-fibre-scalar-commutator}
\end{equation}
Two derivatives then give
\begin{equation}
 [J(u,y),J(u',y')]
 =i\partial_u\delta(u-u')\,\delta^{d-2}(y-y').
 \label{eq:null-fibre-current-commutator}
\end{equation}
These are the two-point function and commutator of a chiral
$\mathrm U(1)$ current, with independent degrees of freedom along the
different null generators. More precisely, the one-particle space
decomposes as a direct integral of current fibres over the transverse
plane \cite{Morinelli:2021null}. In a transverse regulator, this
becomes a collection of independent currents. This is the null
quantization used in
\cite{Wall:2011hj,Bousso:2015wca,Malik:2019dpg,Moosa:2020jwt,Roy:2022yzm}.

We may therefore work on one current fibre and suppress $y$. For the
canonical scalar stress tensor, our normalization gives
\begin{equation}
 T_{++}(u)=\frac12:J(u)^2:,
 \label{eq:null-coherent-current-energy}
\end{equation}
where normal ordering is with respect to the vacuum.

To construct a coherent excitation, choose a real compactly supported test function
$h\in C_c^\infty(\mathbb R)$ with $\int h(u)\,\dd u=0$, and set
\begin{equation}
 \phi(h)\defeq\int_{\mathbb R}\dd u\,h(u)\phi(u),
 \qquad
 W(h)\defeq e^{i\phi(h)},
 \qquad
 |\Omega_h\rangle\defeq W(h)|\Omega\rangle.
 \label{eq:null-coherent-weyl-state}
\end{equation}
The smeared field is self-adjoint and $W(h)$ is unitary. We denote
the resulting state by $\rho_h$, so that
\begin{equation}
 \langle A\rangle_{\rho_h}
 =\langle\Omega|W(h)^\dagger A W(h)|\Omega\rangle.
\end{equation}
The construction extends by completion in the one-particle norm
defined by \eqref{eq:null-fibre-scalar-smearing-norm}. In particular,
the smearing profile used below is admissible even though it has
a $u^{-2}$ tail and is not compactly supported.

The scalar commutator implies
\begin{equation}
 [\phi(h),J(u)]=\frac{i}{\sqrt{2}}h(u).
\end{equation}
Since this is proportional to the identity, the Baker-Campbell-Hausdorff expansion terminates after
the first term and gives
\begin{equation}
 W(h)^\dagger J(u)W(h)=J(u)+\frac{h(u)}{\sqrt{2}}.
 \label{eq:null-coherent-current-shift}
\end{equation}
The excitation therefore has current profile
\begin{equation}
 j_h(u)\defeq\langle J(u)\rangle_{\rho_h}
 =\frac{h(u)}{\sqrt{2}}.
 \label{eq:null-coherent-current-profile}
\end{equation}
Conversely, an admissible real profile $j$ with vanishing integral
specifies the smearing $h_j=\sqrt{2}j$. We write the corresponding
state as $\rho_{\mathrm{coh}}[j]$.

The coherent displacement changes the mean but leaves the centered covariance
equal to that of the vacuum,
\begin{equation}
 \left\langle
  \bigl(J(u)-j(u)\bigr)\bigl(J(u')-j(u')\bigr)
 \right\rangle_{\rho_{\mathrm{coh}}[j]}
 =\langle J(u)J(u')\rangle_\Omega.
 \label{eq:null-coherent-unchanged-covariance}
\end{equation}
The vacuum subtraction in the stress tensor consequently gives
\begin{align}
 \langle T_{++}(u)\rangle_{\rho_{\mathrm{coh}}[j]}
 &=\frac12\lim_{u'\to u}
   \left[
    \langle J(u)J(u')\rangle_{\rho_{\mathrm{coh}}[j]}
    -\langle J(u)J(u')\rangle_\Omega
   \right]
 \notag\\
 &=\frac12j(u)^2.
 \label{eq:null-coherent-current-one-point}
\end{align}
A Gaussian state is determined by its mean and covariance. It follows
that the restriction of this coherent state to $u>V$ depends only on
the profile in that half-line. Different completions in $u<V$ that
enforce the global zero-mode condition can therefore give the same
restricted state.

\subsection{Modular frequencies and regulated states}
\label{subsec:null-coherent-modular-frequencies}

Let $\rho_V$ and $\sigma_V$ denote the restrictions of the coherent
state and the vacuum to $u>V$. By the half-line Bisognano--Wichmann theorem \cite{Bisognano:1975duality}, vacuum modular flow acts by dilations
about the endpoint. It is therefore useful to measure distance from
the cut logarithmically,
\begin{equation}
 x\defeq u-V,
 \qquad
 r\defeq\log\frac{x}{L},
 \qquad x=Le^r,
 \label{eq:null-coherent-log-coordinate}
\end{equation}
where $L>0$ is an arbitrary reference length. Define
\begin{equation}
 \mathcal J_V(r)\defeq xJ(V+x),
 \qquad
 \psi_V(r)\defeq\langle\mathcal J_V(r)\rangle_{\rho_V}
 =xj(V+x).
 \label{eq:null-coherent-mellin-profile}
\end{equation}
The current has scaling dimension one, and the factor $x$ compensates
this scaling weight, so modular dilations act on $\mathcal J_V$ by
translations in $r$. Its Fourier modes consequently diagonalize
modular flow.

The same change of variables simplifies the modular energy. Using
the vacuum modular Hamiltonian and
\eqref{eq:null-coherent-current-one-point}, we find
\begin{align}
 \Delta\langle H[V]\rangle_j
 &=2\pi\int_V^\infty\dd u\,(u-V)
   \langle T_{++}(u)\rangle_{\rho_V}
 \notag\\
 &=\pi\int_0^\infty\dd x\,xj(V+x)^2
 =\pi\int_{\mathbb R}\dd r\,\psi_V(r)^2.
 \label{eq:null-coherent-modular-energy-profile}
\end{align}
Thus finite modular energy is precisely the condition
$\psi_V\in L^2(\mathbb R,\dd r)$ for these coherent profiles.

In the logarithmic coordinate, the current commutator becomes
\begin{equation}
 [\mathcal J_V(r),\mathcal J_V(r')]=i\delta'(r-r').
\end{equation}
We use the corresponding mode expansion
\begin{equation}
 \mathcal J_V(r)
 =\int_0^\infty\dd\nu\,\sqrt{\frac{\nu}{2\pi}}
  \left(b_{\nu,V}e^{-i\nu r}
       +b_{\nu,V}^\dagger e^{i\nu r}\right),
 \qquad
 [b_{\nu,V},b_{\nu',V}^\dagger]=\delta(\nu-\nu').
 \label{eq:null-coherent-current-mode-expansion}
\end{equation}
By the Bisognano--Wichmann theorem, the restricted vacuum is
a KMS state at inverse temperature $2\pi$ for dilations about
the cut \cite{Bisognano:1975duality}. Since these dilations
act as translations of $r$, the frequency $\nu$ corresponds
to modular energy $2\pi\nu$. The KMS condition and the
canonical commutation relations then give
\begin{equation}
 \langle b_{\nu,V}^\dagger b_{\nu',V}\rangle_{\sigma_V}
 =\frac{\delta(\nu-\nu')}{e^{2\pi\nu}-1},
 \qquad
 \langle b_{\nu,V}b_{\nu',V}\rangle_{\sigma_V}=0.
 \label{eq:null-coherent-modular-thermal-correlations}
\end{equation}
The formal oscillator expression for the regional modular Hamiltonian is
\begin{equation}
 H[V]=2\pi\int_0^\infty\dd\nu\,\nu
       b_{\nu,V}^\dagger b_{\nu,V}+\text{constant}.
 \label{eq:null-coherent-modular-mode-hamiltonian}
\end{equation}
Here the continuum modes are spectral notation for the current and
its thermal correlations. Density matrices and the vacuum-centering
constant will be defined on a finite collection of oscillators below.

The coherent state is specified in this basis by its displacement
amplitudes
\begin{equation}
 \gamma_V(\nu)\defeq\langle b_{\nu,V}\rangle_{\rho_V}.
\end{equation}
Taking the expectation value of
\eqref{eq:null-coherent-current-mode-expansion} gives
\begin{equation}
 \psi_V(r)
 =\int_0^\infty\dd\nu\,\sqrt{\frac{\nu}{2\pi}}
  \left(\gamma_V(\nu)e^{-i\nu r}
       +\overline{\gamma_V(\nu)}e^{i\nu r}\right).
 \label{eq:null-coherent-profile-mode-expansion}
\end{equation}
With the Fourier convention
\begin{equation}
 \widehat\psi_V(\omega)
 \defeq\int_{\mathbb R}\dd r\,e^{-i\omega r}\psi_V(r),
 \label{eq:null-coherent-mellin-transform}
\end{equation}
we obtain, for $\nu>0$,
\begin{equation}
 \widehat\psi_V(-\nu)=\sqrt{2\pi\nu}\,\gamma_V(\nu),
 \qquad
 \widehat\psi_V(\nu)=\sqrt{2\pi\nu}\,
                    \overline{\gamma_V(\nu)}.
 \label{eq:null-coherent-profile-mode-relation}
\end{equation}
Parseval's identity then gives the equivalent energy formulas
\begin{equation}
 \Delta\langle H[V]\rangle_j
 =\frac12\int_{\mathbb R}\dd\omega\,
     |\widehat\psi_V(\omega)|^2
 =2\pi\int_0^\infty\dd\nu\,\nu|\gamma_V(\nu)|^2.
 \label{eq:null-coherent-modular-energy-spectrum}
\end{equation}
The last expression controls a weighted norm of $\gamma_V$. It does
not require $\gamma_V\in L^2(\mathbb R_+,\dd\nu)$, which would be
needed to implement the shift by a displacement operator in the
ordinary Fock representation of these modes. We will construct the
displacements at a finite regulator and describe their continuum
limits through current profiles and correlation functions.

For this purpose, retain frequencies in a window
$\epsilon\leq\nu\leq\Lambda$, with $0<\epsilon<\Lambda<\infty$,
and partition the window into finitely many disjoint intervals $I_a$
of widths $w_a$. Define normalized frequency packets by
\begin{equation}
 b_{a,V}\defeq\frac1{\sqrt{w_a}}\int_{I_a}\dd\nu\,b_{\nu,V},
 \qquad
 \gamma_{V,a}\defeq\frac1{\sqrt{w_a}}
                    \int_{I_a}\dd\nu\,\gamma_V(\nu).
 \label{eq:null-coherent-regulated-frequency-packets}
\end{equation}
The disjoint intervals give
$[b_{a,V},b_{a',V}^\dagger]=\delta_{aa'}$. The vacuum occupation
numbers are
\begin{equation}
 n_a\defeq\frac1{w_a}\int_{I_a}
                  \frac{\dd\nu}{e^{2\pi\nu}-1},
 \qquad
 \kappa_a\defeq\log(1+n_a^{-1}).
 \label{eq:null-coherent-regulated-thermal-parameters}
\end{equation}
There are no correlations between distinct packets. Since the vacuum
is Gaussian, its restriction to these oscillators is exactly a product
of thermal states. Suppressing the regulator in the notation and
writing $N_a=b_{a,V}^\dagger b_{a,V}$, we have
\begin{equation}
 \sigma_V=\bigotimes_a\sigma_{\kappa_a},
 \qquad
 \sigma_{\kappa_a}=(1-e^{-\kappa_a})e^{-\kappa_aN_a}.
 \label{eq:null-coherent-regulated-vacuum-factorization}
\end{equation}
The effective spacing $\kappa_a$ approaches $2\pi\nu$ as the packet
narrows around $\nu$. At finite packet width, it is determined by the
averaged occupation number in
\eqref{eq:null-coherent-regulated-thermal-parameters}.

The coherent state has the same centered covariance and the means
$\gamma_{V,a}$. Let
\begin{equation}
 D_a(\gamma)\defeq
 \exp\!\left(\gamma b_{a,V}^\dagger-\overline\gamma b_{a,V}\right),
 \qquad
 D_a(\gamma)^\dagger b_{a,V}D_a(\gamma)=b_{a,V}+\gamma.
 \label{eq:null-coherent-mode-displacement}
\end{equation}
The regulated coherent state is therefore
\begin{align}
 D_V(\boldsymbol\gamma_V)&\defeq\bigotimes_aD_a(\gamma_{V,a}),
 \notag\\
 \rho_V
 &=D_V(\boldsymbol\gamma_V)\sigma_V
   D_V(\boldsymbol\gamma_V)^\dagger
 =\bigotimes_a\rho_{\gamma_{V,a}},
 \notag\\
 \rho_{\gamma_{V,a}}
 &\defeq D_a(\gamma_{V,a})\sigma_{\kappa_a}
          D_a(\gamma_{V,a})^\dagger.
 \label{eq:null-coherent-regulated-state-factorization}
\end{align}
This displacement implements the mean shift induced by $W(h)$ on
the retained oscillator algebra.

The centered modular Hamiltonian of that algebra is
\begin{equation}
 H_{\mathrm{reg}}[V]=\sum_a\kappa_a(N_a-n_a),
 \qquad
 \Tr\!\left(\rho_VH_{\mathrm{reg}}[V]\right)
 =\sum_a\kappa_a|\gamma_{V,a}|^2.
 \label{eq:null-coherent-regulated-modular-energy}
\end{equation}
Refining the partition at fixed $\epsilon,\Lambda$ recovers the
spectral energy integral over that window. The infrared and
ultraviolet cutoffs can then be removed whenever
\eqref{eq:null-coherent-modular-energy-spectrum} is finite.

The regulator keeps finitely many oscillators, each with arbitrarily
high occupation numbers. The regulated states are therefore faithful, with
positive eigenvalues that approach zero. Since the argument in
Section~\ref{sec:az-null-variations} requires the spectrum to be
bounded away from zero, we define the products involving inverse
powers directly using the Gaussian formulas for displaced thermal
oscillators. For the profile studied below, we use frequency estimates to show
that the modular energies and their first two derivatives with
respect to the cut converge as the regulator is removed. This
allows us to evaluate the transport term from the modular-energy
response in \eqref{eq:az-transport-modular-response}.

\subsection{Coherent escort states}
\label{subsec:null-coherent-escorts}

We now construct the fixed-ray escorts of the regulated coherent
state. The calculation applies to every $c>0$ and $\beta>0$. Each
escort will have the same centered covariance as the restricted
vacuum, with a displacement that depends on the modular frequency.

Keep the cut $V$ and the oscillator regulator fixed, and set
\begin{equation}
 a_{\beta,c}\defeq\frac{1-\beta}{2c\beta},
 \qquad
 Y_{\beta,c}[V]\defeq
 \sigma_V^{a_{\beta,c}}\rho_V^{1/c}\sigma_V^{a_{\beta,c}},
 \qquad
 \eta_{\beta,c}[V]\defeq
 \frac{Y_{\beta,c}[V]^{c\beta}}
      {\Tr Y_{\beta,c}[V]^{c\beta}}.
 \label{eq:null-coherent-full-escort}
\end{equation}
The product states in
\eqref{eq:null-coherent-regulated-vacuum-factorization} and
\eqref{eq:null-coherent-regulated-state-factorization} reduce the
calculation to one oscillator. With the positive closures specified
below, functional calculus gives
\begin{align}
 Y_{\beta,c}[V]&=\bigotimes_aY_{\beta,c;a},
 &
 Y_{\beta,c;a}
 &=\sigma_{\kappa_a}^{a_{\beta,c}}
   \rho_{\gamma_{V,a}}^{1/c}
   \sigma_{\kappa_a}^{a_{\beta,c}},
 \notag\\
 \eta_{\beta,c}[V]
 &=\bigotimes_a
   \frac{Y_{\beta,c;a}^{c\beta}}
        {\Tr Y_{\beta,c;a}^{c\beta}}.
 \label{eq:null-coherent-escort-mode-factorization}
\end{align}

Suppress the mode label. We use $[b,b^\dagger]=1$,
$N=b^\dagger b$, and the displacement convention of
\eqref{eq:null-coherent-mode-displacement}, so that
\begin{equation}
 \sigma_\kappa=Z_\kappa^{-1}e^{-\kappa N},
 \qquad
 Z_\kappa=(1-e^{-\kappa})^{-1},
 \qquad
 \rho_\gamma=D(\gamma)\sigma_\kappa D(\gamma)^\dagger.
 \label{eq:null-coherent-thermal-mode}
\end{equation}
Unitary covariance of functional calculus gives
$\rho_\gamma^{1/c}=D(\gamma)\sigma_\kappa^{1/c}D(\gamma)^\dagger$.
Consequently,
\begin{equation}
 Y_{\beta,c}
 =Z_\kappa^{-1/(c\beta)}
  e^{-sN}D(\gamma)e^{-tN}D(\gamma)^\dagger e^{-sN},
 \label{eq:null-coherent-mode-factorization}
\end{equation}
where
\begin{equation}
 s\defeq\frac{\kappa(1-\beta)}{2c\beta},
 \qquad
 t\defeq\frac{\kappa}{c},
 \qquad
 T\defeq2s+t=\frac{\kappa}{c\beta}>0.
 \label{eq:null-coherent-gaussian-parameters}
\end{equation}
The two outer factors are unbounded when $\beta>1$. Their combined
product nevertheless has a bounded positive closure. We establish
this directly, together with the Gaussian identity needed to
normalize the escort.

Introduce the unnormalized coherent vectors
$|w)\defeq e^{wb^\dagger}|0\rangle$, with $w\in\mathbb C$.
Normal ordering of the displacement gives
\begin{equation}
 e^{-sN}|w)=|e^{-s}w),
 \qquad
 D(\gamma)|w)
 =e^{-|\gamma|^2/2-\overline\gamma w}|w+\gamma).
 \label{eq:null-coherent-exponential-vector-actions}
\end{equation}
These identities also hold for negative $s$. All the factors in
\eqref{eq:null-coherent-mode-factorization} preserve the dense space
\begin{equation}
 \mathcal D_{\mathrm{an}}
 \defeq\operatorname{span}
 \left\{P(b^\dagger)e^{wb^\dagger}|0\rangle
       \;\middle|\; P\text{ a polynomial},\ w\in\mathbb C\right\}.
 \label{eq:null-coherent-gaussian-core}
\end{equation}
Writing
\begin{equation}
 \mathcal P_{s,t}(\gamma)
 \defeq e^{-sN}D(\gamma)e^{-tN}D(\gamma)^\dagger e^{-sN},
 \qquad
 A\defeq e^{-s}(1-e^{-t})\gamma,
\end{equation}
successive application of
\eqref{eq:null-coherent-exponential-vector-actions} yields
\begin{equation}
 \mathcal P_{s,t}(\gamma)|w)
 =\exp\!\left[-(1-e^{-t})|\gamma|^2+\overline A w\right]
  |e^{-T}w+A).
 \label{eq:null-coherent-gaussian-product-action}
\end{equation}
The same calculation with a single displaced thermal factor gives
\begin{equation}
 D(\delta)e^{-TN}D(\delta)^\dagger|w)
 =\exp\!\left[-(1-e^{-T})|\delta|^2
              +(1-e^{-T})\overline\delta w\right]
  |e^{-T}w+(1-e^{-T})\delta).
 \label{eq:null-coherent-displaced-thermal-action}
\end{equation}
The coherent vectors in \eqref{eq:null-coherent-gaussian-product-action} and  \eqref{eq:null-coherent-displaced-thermal-action}, and their $w$-dependent prefactors agree if
\begin{equation}
 \delta=\frac{e^{-s}(1-e^{-t})}{1-e^{-T}}\,\gamma
       =\frac{\sinh(t/2)}{\sinh(T/2)}\,\gamma
       =\frac{\sinh(\kappa/2c)}
              {\sinh(\kappa/(2c\beta))}\,\gamma.
 \label{eq:null-coherent-mode-multiplier}
\end{equation}
Comparing the remaining scalar factors therefore proves
\begin{align}
 \mathcal P_{s,t}(\gamma)
 &=C_{s,t}(\gamma)D(\delta)e^{-TN}D(\delta)^\dagger,
 \label{eq:null-coherent-gaussian-product}\\
 C_{s,t}(\gamma)
 &\defeq\exp\!\left[
   (1-e^{-T})|\delta|^2-(1-e^{-t})|\gamma|^2
  \right]>0.
 \label{eq:null-coherent-gaussian-scalar}
\end{align}
The calculation establishes
\eqref{eq:null-coherent-gaussian-product} on
$\mathcal D_{\mathrm{an}}$, which is dense in the oscillator
Hilbert space. The right-hand side is a positive
multiple of a displaced thermal operator. Its eigenvalues are
$C_{s,t}(\gamma)e^{-Tn}$ for $n=0,1,\ldots$. Since $T>0$, these
decrease exponentially, so the operator is bounded and has
finite trace. This gives a well-defined expression for the full product,
including when the individual factors $e^{-sN}$ are unbounded.
We use this bounded extension on the full oscillator Hilbert
space when taking powers and computing traces below.

We can now take the power $c\beta$ by the spectral theorem.
The relation $c\beta T=\kappa$ restores the original thermal spacing,
\begin{equation}
 Y_{\beta,c}^{c\beta}
 =Z_\kappa^{-1}C_{s,t}(\gamma)^{c\beta}
  D(\delta)e^{-\kappa N}D(\delta)^\dagger,
 \qquad
 \Tr Y_{\beta,c}^{c\beta}=C_{s,t}(\gamma)^{c\beta}.
 \label{eq:null-coherent-mode-normalization}
\end{equation}
Thus the normalized single-mode escort is
\begin{equation}
 \eta_{\beta,c}=D(\delta)\sigma_\kappa D(\delta)^\dagger.
 \label{eq:null-coherent-mode-escort}
\end{equation}
Its centered covariance and spectrum are those of $\sigma_\kappa$.
At $\beta=1$, $s=0$, $C_{s,t}=1$, and $\delta=\gamma$, as required
by $\eta_{1,c}=\rho_\gamma$.

Applying the single-oscillator result to each packet gives
\begin{align}
 \eta_{\beta,c}[V]
 &=\bigotimes_a
 \left[
 D_a(\delta_{\beta,c;V,a})\sigma_{\kappa_a}
 D_a(\delta_{\beta,c;V,a})^\dagger
 \right],
 \notag\\
 \delta_{\beta,c;V,a}
 &=\frac{\sinh(\kappa_a/2c)}
         {\sinh(\kappa_a/(2c\beta))}\,\gamma_{V,a}.
 \label{eq:null-coherent-full-mode-multiplier}
\end{align}
Thus each packet has the vacuum thermal state displaced by
$\delta_{\beta,c;V,a}$.

To pass to the continuum, we first narrow the packets within
a fixed frequency interval $\epsilon\leq\nu\leq\Lambda$.
The packet spacing $\kappa_a$ then approaches $2\pi\nu$, giving
\begin{equation}
 \delta_{\beta,c;V}(\nu)
 =\frac{\sinh(\pi\nu/c)}
        {\sinh(\pi\nu/(c\beta))}\,\gamma_V(\nu).
 \label{eq:null-coherent-continuum-mode-multiplier}
\end{equation}
The multiplier is smooth and bounded on this interval, so the
packet displacements converge to this expression. We subsequently
take $\epsilon\downarrow0$ and $\Lambda\uparrow\infty$ under the
convergence conditions discussed below. We also denote the
resulting continuum escort by $\eta_{\beta,c}[V]$.

Let $j_{\beta,c;V}$ denote the resulting escort current profile and
write, as in \eqref{eq:null-coherent-mellin-profile},
\begin{equation}
 j_{\beta,c;V}(u)\defeq
 \langle J(u)\rangle_{\eta_{\beta,c}[V]},
 \qquad
 \psi_{\beta,c;V}(r)\defeq xj_{\beta,c;V}(V+x),
 \qquad x=Le^r.
 \label{eq:null-coherent-escort-profile-definition}
\end{equation}
The relation between displacement amplitudes and Fourier
coefficients in \eqref{eq:null-coherent-profile-mode-relation}
then gives
\begin{equation}
 \widehat\psi_{\beta,c;V}(\omega)
 =\mathfrak m_{\beta,c}(\omega)\widehat\psi_V(\omega),
 \qquad
 \mathfrak m_{\beta,c}(\omega)
 \defeq\frac{\sinh(\pi\omega/c)}
              {\sinh(\pi\omega/(c\beta))}.
 \label{eq:null-coherent-modular-filter}
\end{equation}
This real, even multiplier preserves the reality of the current
profile. Its continuous value at the origin is
$\mathfrak m_{\beta,c}(0)=\beta$, and
$\mathfrak m_{1,c}(\omega)=1$. Changing the reference length $L$
only translates $r$, so it multiplies both sides by the same Fourier
phase and has no effect on the physical profile.

For $\beta>1$, forming the escort amplifies the large-frequency
components of the current, since
\begin{equation}
 \mathfrak m_{\beta,c}(\omega)
 \sim\exp\!\left[
  \frac{\pi|\omega|}{c}\left(1-\frac1\beta\right)
 \right],
 \qquad |\omega|\longrightarrow\infty.
 \label{eq:null-coherent-filter-growth}
\end{equation}
The original Fourier coefficients $\widehat\psi_V(\omega)$
must decay sufficiently rapidly to compensate for this growth.
In particular, finite modular energy of the escort requires
\begin{equation}
 \int_{\mathbb R}\dd\omega\,
 |\widehat\psi_{\beta,c;V}(\omega)|^2
 =
 \int_{\mathbb R}\dd\omega\,
 |\mathfrak m_{\beta,c}(\omega)\widehat\psi_V(\omega)|^2
 <\infty.
 \label{eq:null-coherent-escort-energy-domain}
\end{equation}
Finite modular energy of the original state alone does not
guarantee this condition.

The continuum escort is the Gaussian state with mean current
$j_{\beta,c;V}$ and the vacuum connected two-point function.
We require these data to define a normal state on the half-line
current algebra. The next subsection verifies this requirement
for an explicit current profile and establishes the convergence
of the modular energy and its first two cut derivatives.

Finally, unitary equivalence to the vacuum gives the exact
regulated entropy and relative-entropy identities
\begin{align}
 S(\eta_{\beta,c}[V])-S(\sigma_V)&=0,
 \label{eq:null-coherent-fixed-ray-entropy}\\
 D(\eta_{\beta,c}[V]\Vert\sigma_V)
 &=\Tr\!\left[\eta_{\beta,c}[V]H_{\mathrm{reg}}[V]\right]
 =\sum_a\kappa_a|\delta_{\beta,c;V,a}|^2.
 \label{eq:null-coherent-fixed-ray-regulated-energy}
\end{align}
For an admissible coherent
profile satisfying \eqref{eq:null-coherent-escort-energy-domain}, the
half-line current relative-entropy formula
\cite{Longo:2018entropy,Bostelmann:2020coherent} identifies the
continuum quadratic form with the relative entropy. With our current
normalization, it reads
\begin{align}
 E_{\beta,c}[V]
 &\defeq D(\eta_{\beta,c}[V]\Vert\sigma_V)
 =\Delta\langle H[V]\rangle_{\eta_{\beta,c}[V]}
 \notag\\
 &=\pi\int_0^\infty\dd x\,x\,j_{\beta,c;V}(V+x)^2
 =\frac12\int_{\mathbb R}\dd\omega\,
   |\mathfrak m_{\beta,c}(\omega)\widehat\psi_V(\omega)|^2.
 \label{eq:null-coherent-fixed-ray-energy}
\end{align}

\subsection{A solvable profile and null convexity}
\label{subsec:null-coherent-solvable-profile}

We now choose a current profile whose cut dependence can be
computed explicitly. Throughout, $\alpha>1$ and $(\alpha,z)$
lies in the upper data-processing region
\eqref{eq:az-upper-dpi-region}, with $c=z/\alpha$.

Let $g>0$ and $q\in\mathbb R\setminus\{0\}$. Choose a real,
smooth global current profile that agrees with
\begin{equation}
 j_\rho(u)=\frac{qg}{(u+g)^2}
 \qquad (u>-g/2)
 \label{eq:null-coherent-rational-profile}
\end{equation}
and has zero integral on the full line. This is achieved by
cutting off the expression smoothly in the past and adding a
compensating pulse in $u<-g/2$. The zero integral and the
$u^{-2}$ tail give
$\widehat j_\rho(k)=O(|k|(1+|\log|k||))$ near $k=0$,
while integrability of all derivatives controls large $|k|$.
Thus the scalar one-particle norm in
\eqref{eq:null-fibre-scalar-smearing-norm} is finite.
For $|V|<g/4$, the completion lies outside the half-line and
does not affect the restricted state.

Define
\begin{equation}
 g_V\defeq g+V,
 \qquad
 r_V\defeq\log\frac{g_V}{L},
 \qquad
 f(r)\defeq\frac{e^r}{(1+e^r)^2}
           =\frac{1}{4\cosh^2(r/2)}.
 \label{eq:null-coherent-profile-scales}
\end{equation}
The logarithmic profile
\eqref{eq:null-coherent-mellin-profile} becomes
\begin{equation}
 \psi_V(r)=\frac{qg}{g_V}f(r-r_V).
 \label{eq:null-coherent-rational-mellin-profile}
\end{equation}
Using $t=e^r$ and Euler's beta integral gives
\begin{equation}
 \widehat f(\omega)
 =\int_0^\infty\frac{t^{-i\omega}}{(1+t)^2}\,\dd t
 =\Gamma(1-i\omega)\Gamma(1+i\omega)
 =\frac{\pi\omega}{\sinh(\pi\omega)},
 \label{eq:null-coherent-profile-fourier-integral}
\end{equation}
and hence
\begin{equation}
 \widehat\psi_V(\omega)
 =\frac{qg}{g_V}e^{-i\omega r_V}
   \frac{\pi\omega}{\sinh(\pi\omega)}.
 \label{eq:null-coherent-rational-mellin-transform}
\end{equation}
All quotients at zero frequency are understood by continuity.

For $1\leq\beta\leq\alpha$, set
\begin{equation}
 K_{\beta,c}(\omega)
 \defeq\frac{\pi\omega}{\sinh(\pi\omega)}
       \frac{\sinh(\pi\omega/c)}
            {\sinh(\pi\omega/(c\beta))}.
 \label{eq:null-coherent-profile-spectral-kernel}
\end{equation}
Equations~\eqref{eq:null-coherent-modular-filter}
and~\eqref{eq:null-coherent-fixed-ray-energy} then give
\begin{align}
 \widehat\psi_{\beta,c;V}(\omega)
 &=\frac{qg}{g_V}e^{-i\omega r_V}K_{\beta,c}(\omega),
 \label{eq:null-coherent-az-filtered-transform}\\
 E_{\beta,c}[V]
 &=\frac{q^2g^2}{2g_V^2}
   \int_{\mathbb R}\dd\omega\,K_{\beta,c}(\omega)^2,
 \label{eq:null-coherent-az-escort-energy}
\end{align}
whenever the integral is finite. The phase drops out of the
energy, leaving all cut dependence in the factor $g_V^{-2}$.

To justify the continuum limit, first suppose
\begin{equation}
 z>\alpha-1,
 \qquad
 d_{\alpha,z}\defeq1-\frac{\alpha-1}{z}>0.
 \label{eq:null-coherent-spectral-gap}
\end{equation}
Since $K_{\beta,c}(0)=\beta$, the large-frequency behavior
of the hyperbolic functions gives
\begin{equation}
 |K_{\beta,c}(\omega)|
 \leq C(1+|\omega|)e^{-\pi d_{\alpha,z}|\omega|},
 \qquad 1\leq\beta\leq\alpha.
 \label{eq:null-coherent-uniform-frequency-bound}
\end{equation}
Differentiating \eqref{eq:null-coherent-az-filtered-transform}
also gives
\begin{align}
 \left|\partial_V\widehat\psi_{\beta,c;V}(\omega)\right|
 &\leq C_1(1+|\omega|)^2
       e^{-\pi d_{\alpha,z}|\omega|},
 \notag\\
 \left|\partial_V^2\widehat\psi_{\beta,c;V}(\omega)\right|
 &\leq C_2(1+|\omega|)^3
       e^{-\pi d_{\alpha,z}|\omega|}.
 \label{eq:null-coherent-uniform-cut-bound}
\end{align}
The constants are uniform for $|V|\leq g/4$ and
$1\leq\beta\leq\alpha$. Consequently, the energy integrand
and its first two cut derivatives are bounded by an
integrable multiple of
$(1+|\omega|)^4e^{-2\pi d_{\alpha,z}|\omega|}$.

Finite modular energy also ensures that the escorts are
normal coherent states in the thermal representation,
since
\begin{equation}
 \int_0^\infty\dd\nu\,\tanh(\pi\nu)
 |\delta_{\beta,c;V}(\nu)|^2
 \leq\pi\int_0^\infty\dd\nu\,\nu
 |\delta_{\beta,c;V}(\nu)|^2<\infty.
 \label{eq:null-coherent-normality-bound}
\end{equation}

Choose nested packet partitions and frequency windows
exhausting $(0,\infty)$. At fixed cutoffs, shrinking the
packet widths gives the energy integral and its first two
cut derivatives. The uniform bounds \eqref{eq:null-coherent-uniform-frequency-bound} and \eqref{eq:null-coherent-uniform-cut-bound} then justify
removing the cutoffs and performing the escort average and
the two cut derivatives under the integral by dominated
convergence. These packet algebras generate the half-line
current algebra. The convergence theorem
\cite[Theorem~3.5]{Hiai:2024qve} therefore identifies the
limit of the regulated divergences of the original pair
$\rho_V,\sigma_V$ with $D_{\alpha,z}(\rho_V\Vert\sigma_V)$.

Applying the fixed-ray identity
\eqref{eq:az-fixed-ray-integral} to the packet states and
taking this limit gives
\begin{equation}
 D_{\alpha,z}(\rho_V\Vert\sigma_V)
 =\frac{q^2g^2}{g_V^2}\frac{\alpha}{2(\alpha-1)}
  \int_{\mathbb R}\dd\omega
  \int_1^\alpha\frac{\dd\beta}{\beta^2}
  K_{\beta,c}(\omega)^2.
 \label{eq:interchanged-integral-az-omega}
\end{equation}
The $\beta$ integral follows from
\begin{align}
 \int_1^\alpha\frac{\dd\beta}{\beta^2}
 \frac{\sinh^2a}{\sinh^2(a/\beta)}
 &=\frac{\sinh^2a}{a}
   \left[\coth(a/\alpha)-\coth a\right]
 \notag\\
 &=\frac{\sinh a\,\sinh((1-\alpha^{-1})a)}
         {a\sinh(a/\alpha)},
 \label{eq:null-coherent-az-beta-integral}
\end{align}
with continuous value $\alpha-1$ at $a=0$.
Taking $a=\pi\omega/c$, using evenness, and setting
$s=\pi\omega$ on the positive half-line yields
\begin{align}
 D_{\alpha,z}(\rho_V\Vert\sigma_V)
 &=\frac{q^2g^2}{g_V^2}\,
   \mathcal A^{\mathrm{coh}}_{\alpha,z},
 \label{eq:null-coherent-az-divergence}\\
 \mathcal A^{\mathrm{coh}}_{\alpha,z}
 &\defeq\frac{z}{\pi(\alpha-1)}
 \int_0^\infty\dd s\,
 \frac{s\sinh(\alpha s/z)\sinh((\alpha-1)s/z)}
      {\sinh^2s\,\sinh(s/z)}.
 \label{eq:null-coherent-az-positive-coefficient}
\end{align}

The coefficient is strictly positive wherever it is finite.
Its integrand tends to $\alpha(\alpha-1)/z$ at the origin
and obeys
\begin{equation}
 \frac{s\sinh(\alpha s/z)\sinh((\alpha-1)s/z)}
      {\sinh^2s\,\sinh(s/z)}
 \sim 2s\exp\!\left[-\frac{2(z-\alpha+1)}{z}s\right]
 \qquad (s\to\infty).
 \label{eq:null-coherent-az-uv-asymptotic}
\end{equation}
Thus finiteness holds precisely for $z>\alpha-1$.
Within \eqref{eq:az-upper-dpi-region}, this includes
$\alpha/2\leq z\leq\alpha$ when $1<\alpha<2$, and
$\alpha-1<z\leq\alpha$ when $\alpha\geq2$.
At $z=\alpha-1$, retain the frequency cutoff through the
$\beta$ integral. The positive cutoff integral grows
quadratically, so the same convergence theorem gives
$D_{\alpha,z}=+\infty$. This includes the Petz point
$(\alpha,z)=(2,1)$, where the state does not satisfy the
finite-divergence hypothesis of
Conjecture~\ref{hyp:az-qnec}.

In the finite range,
\begin{equation}
 \frac{\dd^2}{\dd V^2}D_{\alpha,z}(\rho_V\Vert\sigma_V)
 =\frac{6q^2g^2}{g_V^4}\,
  \mathcal A^{\mathrm{coh}}_{\alpha,z}>0.
 \label{eq:null-coherent-az-hessian}
\end{equation}

Two limits check the normalization. On the sandwiched line
$z=\alpha$,
$K_{\beta,1}(\omega)=\pi\omega/\sinh(\pi\omega/\beta)$.
Using
$\int_{\mathbb R}(\pi\omega/\sinh(\pi\omega))^2\dd\omega
=\pi/3$
and performing the escort average gives
\begin{equation}
 \mathcal A^{\mathrm{coh}}_{\alpha,\alpha}
 =\frac{\pi\alpha(\alpha+1)}{12}.
 \label{eq:null-coherent-sandwiched-coefficient}
\end{equation}
As $\alpha\downarrow1$ at fixed $c\in[1/2,1]$, the escort
average reduces to $\beta=1$. The frequency bounds \eqref{eq:null-coherent-uniform-frequency-bound} are
uniform near this limit and give
$\mathcal A^{\mathrm{coh}}_{\alpha,c\alpha}\to\pi/6$.
Hence
\begin{equation}
 D_1(\rho_V\Vert\sigma_V)=\frac{\pi q^2g^2}{6g_V^2},
 \qquad
 \frac{\dd^2}{\dd V^2}D_1(\rho_V\Vert\sigma_V)
 =\frac{\pi q^2g^2}{g_V^4}
 =2\pi\langle T_{++}(V)\rangle_{\rho_V}.
 \label{eq:null-coherent-relative-entropy-check}
\end{equation}

Finally, take the global current to be $F(y)j_\rho(u)$,
where $F$ is real, smooth, and compactly supported, and
choose a smooth cut with $|V(y)|<g/4$ on its support.
The null-plane fibre decomposition
\cite{Morinelli:2021null}, tensor additivity at finite
transverse regulator, and removal of that regulator using
the same convergence theorem give
\begin{equation}
 D_{\alpha,z}[V]
 =q^2g^2\mathcal A^{\mathrm{coh}}_{\alpha,z}
  \int\dd^{d-2}y\,
  \frac{F(y)^2}{[g+V(y)]^2}.
 \label{eq:null-coherent-higher-d-az-divergence}
\end{equation}
Compact support and $g+V(y)>3g/4$ justify differentiation
under the transverse integral. The functional Hessian is
purely diagonal,
\begin{align}
 (D_{\alpha,z})''_{++}[V;y]
 &=\frac{6q^2g^2\mathcal A^{\mathrm{coh}}_{\alpha,z}}
         {[g+V(y)]^4}F(y)^2\geq0,
 \label{eq:null-coherent-higher-d-az-hessian}\\
 D^{(2)}_{\alpha,z;\off}[V;y,y']&=0.
 \label{eq:null-coherent-higher-d-off-diagonal}
\end{align}
This establishes the diagonal $\alpha$--$z$ R\'enyi QNEC
for the chosen coherent family throughout the finite part
of the upper data-processing region.

\subsection{The escort transport term and R\'enyi QNEC}
\label{subsec:null-coherent-transport-term}

We now evaluate the separate terms in
\eqref{eq:null-sandwiched-local-identity} on the sandwiched line
$z=\alpha$, where $c=1$. Write
$\eta_\beta^{\,\mathrm{s}}[V]=\eta_{\beta,1}[V]$ and
$j_{\beta,V}=j_{\beta,1;V}$.
Rescaling the Fourier transform
\eqref{eq:null-coherent-profile-fourier-integral} in
\eqref{eq:null-coherent-az-filtered-transform} gives
\begin{align}
 \psi_{\beta,V}(r)
 &=\frac{qg}{g_V}\,\beta^2f\bigl(\beta(r-r_V)\bigr),
 \label{eq:null-coherent-sandwiched-mellin-profile}\\
 j_{\beta,V}(V+x)
 &=\frac{qg\beta^2}{g_V^2}
   \frac{(x/g_V)^{\beta-1}}
        {[1+(x/g_V)^\beta]^2},
 \qquad x>0.
 \label{eq:null-coherent-escort-profile}
\end{align}
At $\beta=1$, this is the original current. For $\beta>1$,
the current at fixed $u$ depends on the cut. Restricting an
escort constructed at $V$ therefore differs from constructing
a new escort at the later cut.

Substitution in \eqref{eq:null-coherent-fixed-ray-energy}
gives
\begin{equation}
 E_\beta[V]\defeq E_{\beta,1}[V]
 =\frac{\pi q^2g^2\beta^3}{6g_V^2}.
 \label{eq:null-coherent-escort-modular-energy}
\end{equation}
The frequency limits, cut derivatives, and escort averages
used below are justified by
\eqref{eq:null-coherent-uniform-frequency-bound}
and \eqref{eq:null-coherent-uniform-cut-bound}.
We remove the frequency regulator at fixed $x>0$ before
taking the boundary limit.

The vacuum-subtracted average escort entropy vanishes at
every oscillator regulator by
\eqref{eq:null-coherent-fixed-ray-entropy},
\begin{equation}
 \mathcal S_\alpha^{\esc,\mathrm{s}}[V]-S(\sigma_V)=0.
 \label{eq:null-coherent-escort-entropy}
\end{equation}
This subtraction and its cut derivatives therefore also
vanish in the continuum limit.

For the local-energy term, set $\xi=x/g_V$. Since each
coherent escort has
$\langle T_{++}\rangle=\tfrac12j_{\beta,V}^2$, its escort
average gives
\begin{align}
 B_\alpha[V;x]
 &\defeq 2\pi
 \langle T_{++}(V+x)\rangle_{\overline\eta_\alpha^{\,\mathrm{s}}[V]}
 \notag\\
 &=\frac{\pi q^2g^2}{g_V^4}\frac{\alpha}{\alpha-1}
   \int_1^\alpha\dd\beta\,
   \frac{\beta^2\xi^{2\beta-2}}{(1+\xi^\beta)^4}.
 \label{eq:null-coherent-averaged-energy-near-cut}
\end{align}
For $0<\xi<1$, the integrand is bounded by $\alpha^2$
and tends to zero for every $\beta>1$ as $\xi\downarrow0$.
Dominated convergence in $\beta$ therefore gives
\begin{equation}
 2\pi\langle T_{++}(V)\rangle_{\overline\eta_\alpha^{\,\mathrm{s}}[V]}
 \defeq\lim_{x\downarrow0}B_\alpha[V;x]=0,
 \qquad \alpha>1.
 \label{eq:null-coherent-averaged-boundary-energy}
\end{equation}

To calculate transport, introduce a spatial offset $\ell>0$
and write $e_{\beta,V}(u)=\tfrac12j_{\beta,V}(u)^2$.
The modular energy at $V+\ell$ of the escort constructed
at $V$ is
\begin{equation}
 E_{\beta,\ell}[V]
 =2\pi\int_{V+\ell}^\infty
   (u-V-\ell)e_{\beta,V}(u)\,\dd u.
 \label{eq:null-coherent-offset-modular-energy}
\end{equation}
Differentiating twice yields
\begin{equation}
 \begin{split}
 E_{\beta,\ell}''[V]
 ={}&2\pi e_{\beta,V}(V+\ell)\\
 &+2\pi\int_{V+\ell}^\infty\dd u\,
 \left[(u-V-\ell)\partial_V^2e_{\beta,V}(u)
                   -2\partial_Ve_{\beta,V}(u)\right].
 \end{split}
 \label{eq:null-coherent-transport-rearrangement}
\end{equation}
Here the derivatives inside the integral hold $u$ fixed.
The first term is the local energy, while the integral
contains the response of the escort state.

The offset can be removed after differentiation. At fixed
$x=u-V$, the profile
\eqref{eq:null-coherent-escort-profile} and its first two
cut derivatives are bounded near $x=0$ and are $O(x^{-2})$
at infinity, uniformly in $\beta$. These bounds imply
convergence of $E_{\beta,\ell}$ and its first two cut
derivatives to those of $E_\beta$, including after the
escort average.

Writing
$\mathcal M_\alpha^{\,\mathrm{s}}[V]
=\int_1^\alpha E_\beta[V]\,\dd\mu_\alpha(\beta)$, we obtain
\begin{equation}
 (\mathcal M_\alpha^{\,\mathrm{s}})''[V]
 =\frac{\alpha}{\alpha-1}
  \int_1^\alpha\frac{\dd\beta}{\beta^2}
  \frac{\pi q^2g^2\beta^3}{g_V^4}
 =\frac{\pi q^2g^2\alpha(\alpha+1)}{2g_V^4}.
 \label{eq:null-coherent-averaged-modular-response}
\end{equation}
Averaging \eqref{eq:null-coherent-transport-rearrangement}
and taking $\ell\downarrow0$ thus gives
\begin{equation}
 \mathfrak X_{\alpha,++}^{\,\mathrm{s}}[V]
 =(\mathcal M_\alpha^{\,\mathrm{s}})''[V]
  -\lim_{x\downarrow0}B_\alpha[V;x]
 =\frac{\pi q^2g^2\alpha(\alpha+1)}{2g_V^4}.
 \label{eq:null-coherent-explicit-transport}
\end{equation}
Since the entropy subtraction and the boundary energy
vanish, \eqref{eq:null-sandwiched-local-identity} reduces to
\begin{equation}
 D_\alpha''[V]
 =\mathfrak X_{\alpha,++}^{\,\mathrm{s}}[V]>0.
 \label{eq:null-coherent-hessian-is-transport}
\end{equation}
The positive Hessian therefore comes entirely from the
change of the escort state as the cut moves.

For the transverse profile $F(y)$ of
Section~\ref{subsec:null-coherent-solvable-profile},
the same calculation on each fibre gives
\begin{equation}
 \mathfrak X_{\alpha,++}^{\,\mathrm{s}}[V;y]
 =\frac{\pi q^2g^2\alpha(\alpha+1)}
        {2[g+V(y)]^4}\,F(y)^2
 =(D_\alpha)''_{++}[V;y].
 \label{eq:null-coherent-higher-d-transport}
\end{equation}

The order-one limit requires care. At fixed $x>0$,
\eqref{eq:null-coherent-averaged-energy-near-cut} tends
to $2\pi$ times the energy density of the original state,
$\pi q^2g^2/(g_V+x)^4$. Consequently,
\begin{align}
 \lim_{\alpha\downarrow1}\lim_{x\downarrow0}B_\alpha[V;x]
 &=0,
 \notag\\
 \lim_{x\downarrow0}\lim_{\alpha\downarrow1}B_\alpha[V;x]
 &=\frac{\pi q^2g^2}{g_V^4}.
 \label{eq:null-coherent-noncommuting-boundary-limits}
\end{align}
At $\alpha=1$, the escort is the restriction of the fixed
global state $\rho$, so transport vanishes and
$D_1''[V]=2\pi\langle T_{++}(V)\rangle_{\rho_V}$.
The separate local-energy and transport terms thus have
nonuniform order-one limits, although their sum tends
smoothly to the ordinary QNEC result
\eqref{eq:null-coherent-relative-entropy-check}.

Away from the sandwiched line, the two terms need not
have separate finite limits. Take $c<1$ in the upper
data-processing region, with $z>\alpha-1$ so that
\eqref{eq:null-coherent-az-hessian} is finite.
For $1<\beta<\min\{\alpha,1/c\}$, the pole at
$\omega=-ic\beta$ in
\eqref{eq:null-coherent-profile-spectral-kernel}
determines the current near the cut,
\begin{equation}
 j_{\beta,c;V}(V+x)
 \sim\frac{qg}{g_V^2}
      \frac{(c\beta)^2\sin(\pi\beta)}
           {\sin(\pi c\beta)}
      \left(\frac{x}{g_V}\right)^{c\beta-1}.
 \label{eq:null-coherent-general-ray-boundary-profile}
\end{equation}
The coefficient is nonzero and $c\beta<1$, so the local
energy diverges. This occurs on an interval of escort
orders of positive measure, and hence the averaged
local energy also diverges. The modular energy remains
finite because its integrand behaves as $x^{2c\beta-1}$.

The cancellation is controlled by using the same offset
$\ell$ in both terms of
\eqref{eq:null-coherent-transport-rearrangement}.
For any $0<p<c$, shifting the Fourier contour within
the pole-free strip $|\operatorname{Im}\omega|<c$
bounds the current and its first two cut derivatives
at fixed $x$ by $O(x^{p-1})$ near zero and
$O(x^{-p-1})$ at infinity, uniformly in $\beta$.
Their squares are integrable with the modular weight
$x$, which justifies removing the offset from the
modular energy and its first two derivatives.
The averaged local-energy term then tends to $+\infty$
and transport to $-\infty$, while their sum approaches
the finite Hessian \eqref{eq:null-coherent-az-hessian}.
For $c<1$, the identity \eqref{eq:az-local-free-energy}
must therefore be evaluated by combining these terms
before removing the spatial cutoff.

\subsection{Beyond coherent states}
\label{subsec:null-beyond-coherent-states}

The coherent-state calculation shows that escort transport can
supply the entire positive null Hessian. Two distinct properties
make this example particularly simple. First, each coherent
escort is a displacement of the restricted vacuum. It therefore
has the vacuum covariance and, at the same regulator, the same entropy.
This explains the vanishing entropy subtraction in
\eqref{eq:null-coherent-escort-entropy}. This property does not
depend on the particular current profile, provided the escort
construction is well defined.

The vanishing local energy at the cut is a further property of
the chosen profile. On the sandwiched line,
\eqref{eq:null-coherent-escort-profile} behaves as
$x^{\beta-1}$ near the cut for $\beta>1$, and its escort average
has the zero boundary energy found in
\eqref{eq:null-coherent-averaged-boundary-energy}.
Coherence alone does not imply this behavior. The simple cut
dependence in \eqref{eq:null-coherent-az-divergence} and the
finiteness condition $z>\alpha-1$ also rely on our choice of
profile.

Squeezed states would allow us to study an entropy contribution
alongside local energy and transport. These states change the field
fluctuations while retaining a Gaussian description, so their
escort entropies need not equal the vacuum entropy. To apply
our construction, one would choose a fixed global squeezed
state and restrict its two-point function to each region
$R[V]$. At a finite Gaussian regulator, this determines the
reduced density matrix, and Gaussian multiplication gives its
escort \cite{Seshadreesan:2018renyi}. The escort covariance
then determines its entropy and stress tensor. Their dependence
on the cut supplies the entropy variation and the modular-energy
response in \eqref{eq:az-transport-modular-response}.

Additional regularity conditions would be needed in the continuum.
A natural starting point is a state whose two-point function
differs smoothly from the vacuum two-point function, so that
the vacuum-subtracted local stress tensor is well defined.
This alone does not ensure that the inverse-power products
defining the escorts have finite normalizing traces. That
condition must be checked throughout $1\leq\beta\leq\alpha$
and for nearby cuts.

One must also control the limits of the vacuum-subtracted
entropies and modular energies, including their first two cut
derivatives. Bounds uniform in the escort order and the cut,
analogous to \eqref{eq:null-coherent-uniform-frequency-bound}
and \eqref{eq:null-coherent-uniform-cut-bound}, would justify
interchanging differentiation, escort averaging, and regulator
removal. Boundary evaluation requires a further check.
As the preceding subsection illustrates, a finite total
Hessian need not imply separately finite local-energy and
transport terms.

We leave this squeezed-state calculation beyond the present
scope. For the sandwiched divergence, null convexity for
$\alpha>1$ is already established beyond coherent states in
the relevant free-field theories in $d>2$
\cite{Moosa:2020jwt,Roy:2022yzm}. The purpose of this extension
would be to understand how entropy, local energy, and transport
combine when the simplifications of the coherent example
are absent.

\section{Summary and future directions}\label{sec:conclusions}
In this paper, we have developed quantum-field-theoretic consequences of the
fixed-ray escort representation of $\alpha$--$z$ R\'enyi divergence.
The underlying operator-algebraic identity, established in the companion
work \cite{Kibe:2026dtz}, holds for normal states on arbitrary von Neumann
algebras and is therefore meaningful for Type~III local algebras.

Within the regulated density-matrix framework used in this paper, the representation yields a modular-energy--entropy decomposition, an escort-averaged entanglement first law,
and an escort average of BKM metrics.  For vacuum balls it gives two Bekenstein-type bounds.
For null-deformed regions it yields the complete bilocal shape Hessian and
isolates the stress-tensor contact term.  Assuming diagonal null convexity
in the upper data-processing region, we obtained a quasi-local
$\alpha$--$z$ QNEC involving the null energy of the escort average state and
an escort-transport correction.  Setting $c=1$, equivalently $z=\alpha$,
recovers the corresponding R\'enyi QNEC bound. For coherent state excitations in a free scalar field theory, we computed each term in the quasi-local R\'enyi QNEC and showed that the $\alpha$--$z$ QNEC is satisfied for this family of states.

We discuss several directions for future research below. Operational
formulas involving the $\alpha$--$z$ integral representation and applications to the sandwiched strong-converse exponent for state discrimination are described in the companion work \cite{Kibe:2026dtz}.

\subsection*{Further QFT tests of the $\alpha$--$z$ QNEC}

The coherent-current family studied in Section~\ref{sec:null-coherent-transport} provides a first explicit check of the diagonal $\alpha$--$z$ QNEC away from the sandwiched slice, throughout the part of the upper data-processing region in which the divergence is finite.  
This family is nevertheless highly special, since it is obtained by a Weyl displacement of the vacuum, has unchanged Gaussian covariance, and factorizes over transverse null fibres.  
Away from the sandwiched line, an important next step is to test diagonal null convexity for more general states, including squeezed and mixed Gaussian states, finite-particle excitations, and states with nontrivial transverse correlations.
The latter would also provide genuinely nontrivial tests of the off-diagonal Hessian.

More dynamical and non-Gaussian examples could include local and global quenches \cite{Kibe:2025cqc}, local-operator excitations, and other nonequilibrium states in two-dimensional CFT.  Conformal and replica methods \cite{Lashkari:2014yva} may make the corresponding $\alpha$--$z$ divergences accessible in these settings.  Such calculations could test stability under variation of the ray slope, clarify the behavior on the Petz segment and near the boundary of the data-processing region, and reveal whether counterexamples occur for less constrained states.  Modular-theoretic calculations of Petz--R\'enyi relative entropy in free QFT \cite{Frob:2024ijk} provide another possible starting point.

\subsection*{Quasi-local terms beyond the coherent family}

Beyond testing positivity of the total Hessian, it is important to resolve its separate quasi-local contributions in less special states.  The present coherent calculation achieves this completely on the sandwiched slice, where the local-energy and entropy terms vanish and the Hessian is carried by escort transport.  General free-field states and quench or operator-insertion states in two-dimensional CFT provide natural settings in which to construct the moving escorts and determine their escort-average, Holevo, and transport contributions.  These calculations could identify when escort formation commutes with restriction and whether the individual correction terms possess useful sign properties.

\subsection*{Holography and semiclassical gravity}

Our encoding result gives an exact relation between code and
boundary escort states. In finite-dimensional operator-algebra
quantum error-correcting codes with exact complementary recovery,
escort formation commutes with the reduced encoding channel,
with both the excited state and the reference state encoded by
the same channel. Encoding also preserves the relative entropy
of each escort with respect to the reference state. This extends
the bulk--boundary equality of sandwiched divergence
\cite{Caginalp:2022uzd} to a statement about the individual
escort states. Such codes provide models of subregion duality
and bulk reconstruction
\cite{harlow2018tasi,Jahn:2021uqr,Chen:2021lnq,Kibe:2021gtw}.

A semiclassical interpretation requires additional input.
Holographic studies of the refined sandwiched divergence
\cite{Bao:2019aol} and of R\'enyi divergences in JT gravity
\cite{Ugajin:2020dyd} provide useful starting points. To interpret
the fixed-ray integral in the bulk, however, one needs compatible
bulk representatives for every escort order $\beta\in I_\alpha$
within a common semiclassical code subspace. This requirement
applies also on the sandwiched line. The replica preparations
at integer escort orders in \cite{Bao:2019aol} do not by themselves
establish a bulk description of the continuous average.

If such representatives exist and the bulk--boundary
relative-entropy relation applies to them, the integral
representation would express $D_{\alpha,z}$ as an average of
bulk relative entropies. At quadratic order about the vacuum,
the corresponding statement would involve an average of bulk
canonical energies, as discussed in
Section~\ref{subsec:az-holographic-outlook}. Neither interpretation
requires the escort average itself to describe a single geometry.
For $c\neq1$, constructing bulk representatives of the escorts
defined by the two $c$-dependent power maps remains an open problem.

Approximate quantum error-correcting codes \cite{Cotler:2017erl}
raise a further question. One can compare the state obtained
by encoding an escort with the escort formed from the encoded
states. It would be useful to determine whether their difference
can be bounded in terms of the recovery error and used to control
corrections to bulk--boundary equality of the divergences.

\begin{acknowledgments}
T.K. is supported by a Simons Foundation fellowship through the Targeted Grant to Instituto Balseiro. The work of P.R. has been supported by the Polish National Science Centre through Sonata grant (2022/47/D/ST2/02058). 
During the preparation of this manuscript, the authors used Anthropic Claude and OpenAI ChatGPT as editorial aids in drafting and revising portions of the text. Every mathematical and physical statement, calculation, and citation in the final manuscript was independently checked by the authors. The authors take full responsibility for every aspect of this work.
\end{acknowledgments}

\begin{appendix}

\section{Fr\'echet derivatives}\label{app:frechet-derivatives}

We collect the finite-dimensional Fr\'echet-derivative formulas used in
the regulated density-matrix calculations of the main text.  Their
application to QFT is understood under the regularity assumptions stated
in Section~\ref{sec:qft}. Our conventions follow \cite{hiai2014introduction}.  Let
$\mathbb M_n^{\mathrm{sa}}(a,b)$ denote the set of self-adjoint $n\times n$
matrices whose spectra are contained in the open interval
$(a,b)\subset\mathbb R$.

\begin{definition}[Divided differences]
For a scalar function $f:(a,b)\to\mathbb R$, define
\begin{equation}
 f^{[0]}(x_1)\defeq f(x_1),
 \qquad
 f^{[1]}(x_1,x_2)
 \defeq\frac{f(x_1)-f(x_2)}{x_1-x_2}
 \quad (x_1\neq x_2).
\end{equation}
Recursively, for pairwise distinct arguments, set
\begin{equation}
 f^{[m]}(x_1,\ldots,x_{m+1})
 \defeq
 \frac{
 f^{[m-1]}(x_1,\ldots,x_m)
 -f^{[m-1]}(x_2,\ldots,x_{m+1})}
 {x_1-x_{m+1}}.
\end{equation}
If $f\in C^m(a,b)$, this function has a unique continuous extension to
coincident arguments.  In particular,
\begin{equation}
 f^{[m]}(x,\ldots,x)=\frac{f^{(m)}(x)}{m!}.
\end{equation}
\end{definition}

\begin{definition}[Fr\'echet derivative]
Let $F:\mathbb M_m\to\mathbb M_n$ be defined in a neighborhood of
$A\in\mathbb M_m$.  Its Fr\'echet derivative at $A$ is the linear map
$D F_A:\mathbb M_m\to\mathbb M_n$ satisfying
\begin{equation}
 \frac{\|F(A+X)-F(A)-D F_A[X]\|_2}{\|X\|_2}
 \longrightarrow0
 \qquad\text{as }X\longrightarrow0,
\end{equation}
where $\|X\|_2=(\Tr X^\dagger X)^{1/2}$ is the Hilbert--Schmidt norm.
When $D F$ is itself Fr\'echet differentiable, the second derivative
$D^2F_A[X,Y]$ is a bilinear map.  For a one-parameter path
$A(\varepsilon)$,
\begin{equation}
 \left.\frac{\dd^2}{\dd\varepsilon^2}F(A(\varepsilon))
 \right|_{\varepsilon=0}
 =D F_{A(0)}[A''(0)]
 +D^2F_{A(0)}[A'(0),A'(0)].
\end{equation}
\end{definition}

The following spectral formulas are used repeatedly; see, for example,
Theorem~3.25 and Example~3.34 of \cite{hiai2014introduction}.

\begin{theorem}[Fr\'echet derivatives of matrix functions]
Let $A=\sum_i\lambda_iP_i\in\mathbb M_n^{\mathrm{sa}}(a,b)$ be its spectral
decomposition, where the $P_i$ are mutually orthogonal spectral
projections and $\sum_iP_i=\one$.

If $f\in C^1(a,b)$, then
\begin{equation}
 D(f)_A[X]
 =\sum_{i,j}
 f^{[1]}(\lambda_i,\lambda_j)P_iXP_j.
 \label{eq:app-first-frechet}
\end{equation}
If $f\in C^2(a,b)$, then
\begin{equation}
 \begin{aligned}
 D^2(f)_A[X,Y]
 =\sum_{i,j,k}
 f^{[2]}(\lambda_i,\lambda_j,\lambda_k)
 \bigl(
 P_iXP_jYP_k+P_iYP_jXP_k
 \bigr).
 \end{aligned}
 \label{eq:app-second-frechet}
\end{equation}
\end{theorem}

\subsection{Derivatives for response kernels}\label{ssec:escort-response-explicit}

The response kernels in \eqref{eq:az-escort-responses} can be calculated
directly by Fr\'echet differentiation.  For fixed $\beta$, abbreviate
\begin{equation}
 a=\frac{1-\beta}{2c\beta},
 \quad A=\sigma^a,
 \quad R=\rho^{1/c},
 \quad B=ARA,
 \quad N=B^{c\beta},
 \quad q=\Tr N,
 \quad \eta=N/q.
 \label{eq:az-response-abbreviations}
\end{equation}
For generic shape directions $i,j$, denote the first order shape variations of the density matrices by $\sigma_i, \sigma_j,\rho_i, \rho_j$ and the second-order variations by $\sigma_{ij}, \rho_{ij}$. The corresponding variations of $A,R,B,N$ are denoted by similar $i,j$ subscripts. The first variations are
\begin{align}
 A_i&=D(t^a)_\sigma[\sigma_i],
 &R_i&=D(t^{1/c})_\rho[\rho_i],
 \notag\\
 B_i&=A_iRA+AR_iA+ARA_i,
 &N_i&=D(t^{c\beta})_B[B_i],
 \label{eq:az-first-response-dictionary}
\end{align}
and
\begin{equation}
 \eta_i=\frac{N_i}{q}-\frac{Nq_i}{q^2},
 \qquad q_i=\Tr N_i.
 \label{eq:az-normalized-first-response}
\end{equation}
At second order,
\begin{align}
 A_{ij}
 &=D(t^a)_\sigma[\sigma_{ij}]
 +D^2(t^a)_\sigma[\sigma_i,\sigma_j],
 \notag\\
 R_{ij}
 &=D(t^{1/c})_\rho[\rho_{ij}]
 +D^2(t^{1/c})_\rho[\rho_i,\rho_j],
 \label{eq:az-second-power-responses}
\end{align}
while
\begin{align}
 B_{ij}={}&A_{ij}RA+AR_{ij}A+ARA_{ij}
 \notag\\
 &+A_iR_jA+A_jR_iA+A_iRA_j+A_jRA_i
 \notag\\
 &+AR_iA_j+AR_jA_i,
 \label{eq:az-second-factor-response}
 \\
 N_{ij}={}&D(t^{c\beta})_B[B_{ij}]
 +D^2(t^{c\beta})_B[B_i,B_j],
 \label{eq:az-second-numerator-response}
\end{align}
and
\begin{equation}
 \eta_{ij}
 =\frac{N_{ij}}q
 -\frac{N_iq_j+N_jq_i+Nq_{ij}}{q^2}
 +\frac{2Nq_iq_j}{q^3},
 \qquad q_{ij}=\Tr N_{ij}.
 \label{eq:az-normalized-second-response}
\end{equation}

On the sandwiched slice $c=1$, the response kernels
\eqref{eq:az-response-abbreviations}--\eqref{eq:az-normalized-second-response} become
\begin{equation}
 R=\rho,
 \qquad
 R_i=\rho_i,
 \qquad
 R_{ij}=\rho_{ij}.
\end{equation}
Thus the inner power map $t^{1/c}$ becomes the identity and its second
Fr\'echet derivative vanishes; the outer power map becomes $t^\beta$.

\end{appendix}

\bibliographystyle{JHEP}
\bibliography{refs}
\end{document}